\documentclass[fleqn,usenatbib]{mnras}

\usepackage{newtxtext,newtxmath}

\usepackage[T1]{fontenc}
\usepackage{ae,aecompl}

\usepackage{graphicx}	
\usepackage{amsmath}	
\usepackage{amssymb}	
\usepackage{epsfig,hyperref}
\usepackage{natbib}

\usepackage{mathtools}
\DeclarePairedDelimiter{\ceil}{\lceil}{\rceil}
\DeclareMathOperator\erf{erf}
\newcommand{\bigO}{\mathcal{O}} 
\newcommand{\etabs}{\ensuremath{\eta_{\mathrm{GBS}}}}

\newcommand{\mstar}{MSTAR} 
\newcommand{\kschain}{KS-CHAIN}
\newcommand{\archain}{AR-CHAIN}
\newcommand{\ketju}{KETJU}
\newcommand{\gadget}{GADGET-3}
\newcommand{\MST}{MST} 

\newcommand{\norm}[1]{\|#1\|}
\newcommand{\abs}[1]{\left|#1\right|}
\newcommand{\ud}{\mathrm{d}}

\DeclareMathOperator{\sgn}{sign}
\newcommand{\vect}[1]{\boldsymbol{#1}}

\newcommand{\derfrac}[2]{\frac{\ud #1}{\ud #2}}

\title[A fast regularised integrator]{MSTAR -- a fast parallelised algorithmically regularised integrator with minimum spanning tree coordinates}

\author[A. Rantala et al.]{Antti Rantala$^{1,2}$\thanks{E-mail: anttiran@mpa-garching.mpg.de}, Pauli Pihajoki$^{2}$, Matias Mannerkoski$^{2}$, Peter H. Johansson$^{2}$, \newauthor Thorsten Naab$^{1}$\\
$^{1}$Max-Planck-Institut f\"ur Astrophysik, Karl-Schwarzchild-Str. 1, 
D-85748, Garching, Germany\\
$^{2}$Department of Physics, Gustaf H\"{a}llstr\"{o}min katu 2, University of Helsinki, Finland
}

\date{Accepted XXX. Received YYY; in original form ZZZ}

\pubyear{2019}

\begin{document}
\label{firstpage}
\pagerange{\pageref{firstpage}--\pageref{lastpage}}
\maketitle

\begin{abstract}
We present the novel algorithmically regularised integration method \mstar{} for high accuracy ($|\Delta E/E| \gtrsim 10^{-14}$) integrations of N-body systems using minimum spanning tree coordinates. The two-fold parallelisation of the $\bigO(N_\mathrm{part}^2)$ force loops and the substep divisions of the extrapolation method allows for a parallel scaling up to $N_\mathrm{CPU}  = 0.2 \times N_\mathrm{part}$. The efficient parallel scaling of \mstar{} makes the accurate integration of much larger particle numbers possible compared to the traditional algorithmic regularisation chain (\archain) methods, e.g. $N_\mathrm{part} = 5000$ particles on $400$ CPUs for $1$ Gyr in a few weeks of wall-clock time. We present applications of \mstar{} on few particle systems, studying the Kozai mechanism and N-body systems like star clusters with up to $N_\mathrm{part} =10^4$ particles. Combined with a tree or a fast multipole based integrator the high performance of \mstar{} removes a major computational bottleneck in simulations with regularised subsystems. It will enable the next generation galactic-scale simulations with up to $10^9$ stellar particles (e.g. $m_\star = 100 M_\odot$ for a $M_\star = 10^{11} M_\odot$ galaxy) including accurate collisional dynamics in the vicinity of nuclear supermassive black holes. 
\end{abstract}



\section{Introduction}

Galactic nuclei and their nuclear stellar clusters are among the densest stellar systems in the entire Universe \citep{Misgeld2011}. The nuclei of massive galaxies also host supermassive black holes (SMBHs) with typical masses in the range of $M=10^{6} \text{--} 10^{10} M_{\odot}$
\citep{Kormendy1995,Ferrarese2005,Kormendy2013}, forming a complex, collisionally evolving stellar-dynamical environment (e.g. \citealt{Merritt2013,Alexander2017}). 
In the $\Lambda$CDM hierarchical picture of structure formation, galaxies grow through mergers and gas accretion, resulting 
in situations where the collisional evolution of a galactic nuclei is intermittently interrupted and transformed by a merger 
(e.g. \citealt{White1978,Begelman1980}). For gas-poor mergers, the more concentrated nucleus with a steeper stellar cusp, will determine the structure of the remnant nucleus immediately after the merger \citep{Holley-Bockelmann1999, Boylan-Kolchin2004}. 

If both of the merging galaxies host central SMBHs, the SMBHs will merge through a three-stage process \citep{Begelman1980}. First on larger scales the SMBHs are brought together through dynamical friction from stars and gas until a hard binary 
is formed with a semi-major axis of $a\sim 10 \ \rm pc$. In the second phase the hard binary will interact with the 
stars in the centre of the merger remnant \citep{Begelman1980, Milosavljevic2001, Milosavljevic2003, Khan2011}, scouring a low-density stellar core in the process (e.g. \citealt{Merritt2006, Lauer2007, Rantala2018, Rantala2019}). The largest uncertainty in this process is the rate at which the `loss cone' is depleted, but there is an emerging consensus that the binary will avoid the so called final-parsec problem and eventually merge into a single SMBH, even in the collisionless limit (e.g. \citealt{Berczik2006, Vasiliev2015, Gualandris2017, Ryu2018,Mannerkoski2019}), with the final coalescence driven by the emission of gravitational waves \citep{Peters1963}.

If the galaxy merger is gas-rich, the evolution of the nucleus of the merger remnant proceeds very differently. During galaxy mergers the resulting asymmetric potential effectively funnels gas inwards into the central regions causing a central starburst which rapidly increases the nuclear stellar density by several orders of magnitude (e.g. \citealt{Sanders1996}). In addition, the gas also plays an important role by causing additional drag on the SMBHs \citep{Beckmann2018}, as well as  by forming circumbinary disks which can have a significant and complicated effects on the angular momentum evolution of the binary \citep{Tang2017,Moody2019,Duffell2019}. In general, SMBH mergers are thought to occur very rapidly in dense gas-rich environments (e.g. \citealt{Khan2016}) mainly due to the short duration of the stellar interaction phase of the binary evolution \citep{Quinlan1996}.

The growth of SMBHs and the formation of binaries in galaxy mergers have been extensively studied in recent decades.  
A typical numerical approach is to use grid codes \citep{Kim2011,Dubois2013,Hayward2014}, smoothed particle hydrodynamics codes with tree gravity \citep{Springel2005, Mayer2007, Johansson2009} or direct summation codes \citep{Berczik2006,Khan2011}. The drawback of grid codes and softened tree codes is that they cannot properly treat collisional systems as the grid cell size or the employed gravitational softening length places a strict spatial resolution limit on the simulation. 

Direct summation codes on the other hand are very well suited for studying collisional stellar systems with $N_\mathrm{part} \lesssim 10^6$ particles, such as globular clusters. However, the steep $\bigO(N_\mathrm{part}^2)$ scaling of the direct summation algorithm limits the applicability of this method to systems with much higher particle numbers. In addition, the most widely used direct summation codes are rarely coupled with a hydrodynamic solver. One possibility is to use on-the-fly code switching \citep{Milosavljevic2001,Khan2016}, but this type of procedure is typically cumbersome and may introduce spurious numerical effects into the simulation. We thus argue that a self-consistent numerical framework for simulating SMBH dynamics in a realistic collisional environment with a high stellar density and including also a gas component still remains to be developed. One of the major obstacles for developing such a code has been the lack of available fast accurate small-scale collisional post-Newtonian integrators, which are also straightforward to couple to both large-scale gravity solvers and hydrodynamical methods.

The most widely used method to treat binaries and close encounters of simulation particles in the collisional regions of simulations is the technique of regularisation. The key idea of regularisation is to transform the equations of motion of a dynamical system into a form without coordinate singularities, which makes solving the dynamics significantly easier using standard numerical integration techniques \citep{Aarseth2003}. The first such method with practical numerical applications was the two-body Kustaanheimo-Stiefel (KS) regularisation method \citep{Kustaanheimo1965}, which transformed both the time and spatial coordinates of the system.  
A major step forward for regularised dynamics was the introduction of the chain concept. By arranging the simulation particles into a chain of inter-particle vectors, the \kschain{} of \cite{Mikkola1993} reduced the number of required KS transformations from $N_\mathrm{part}(N_\mathrm{part}-1)/2$ to $N_\mathrm{part}-1$ yielding a much more efficient regularisation method. In the N-body codes of \cite{Aarseth1999} the \kschain{} treats the mutual encounters of not more than six particles simultaneously.

A new regularisation method which does not require a coordinate transformation but only a time transformation was discovered by \cite{Mikkola1999a} and \cite{Preto1999}. This algorithmic regularisation (AR) method is faster than the previous regularisation methods and more accurate, especially in the case of large mass ratios between particles in an N-body system. Many current direct summation codes \citep{Harfst2008,Aarseth2012} and regularised tree codes \citep{Rantala2017} use an implementation of the \archain{} integrator \citep{Mikkola2006,Mikkola2008} to resolve the small-scale dynamics around SMBHs.
 
Despite the many successes of regularisation methods, their original design as few-body codes still limits their applicability to systems with a very large number of particles, which is crucial for performing galactic-scale simulations at high accuracy. 
The regularised tree code \ketju{} \citep{Rantala2017} is currently limited to a particle resolution of $N_\mathrm{part} \lesssim 10^7$ stellar particles per galaxy, although 
the underlying tree code \gadget{} \citep{Springel2005} could easily run simulations with $\sim 10 \text{--} 100$ times more stellar particles, up to $N_\mathrm{part} \sim 10^9$ collisionless particles in a single galaxy. This is because the included \archain{} integrator becomes impractically inefficient with more than $\sim 300$--$500$ particles in the regularised chain region. We note that similar performance issues with the \kschain{} algorithm have already been reported in the literature, see e.g. \cite{Milosavljevic2001}.

In this paper we present a new algorithmically regularised integrator \mstar{} developed and implemented with the aim of addressing some of
the shortcomings of the previous algorithms. Our new code contains two main improvements compared to existing algorithmically regularised (AR) integrators. Firstly, We use a minimum spanning tree (\MST{}) coordinate system instead of the chain structure, motivating the name of the code. We note that regularised integration algorithms using particular tree structures have been implemented before: \cite{Jernigan1989} developed a KS-regularised binary tree code while \cite{Mikkola1989} proposed a `branching' KS-chain structure for few-body regularisation. However, neither of these methods are widely used today. Secondly, a major speed-up compared to the previous regularisation methods is achieved by applying a two-fold parallalisation strategy to the extrapolation method, which is used in regularised integrators to ensure a high numerical accuracy \citep{Mikkola1999b}.

The remainder of this paper is organised as follows: in Section \ref{Section: AR-CHAIN} we review our implementation of \archain{}, as the new code \mstar{} was developed based on our earlier experience with the \archain{} integrator. The numerical and parallelisation procedures of the \mstar{} integrator and the code implementation are discussed in Section \ref{Section: MST}.  We describe and test the parallel extrapolation method in Section \ref{Section: parallel}. In Section \ref{Section: code accuracy} we perform a number of few-body code tests to validate the accuracy of the \mstar{} code, whereas in Section \ref{Section: code scaling} we perform a number of scaling and timing tests. Finally, we summarise our results and present our conclusions in Section \ref{Section: conclusions}.

\section{AR-CHAIN} \label{Section: AR-CHAIN}
\subsection{Time transformation of equations of motion}

The algorithmic regularisation chain (\archain) integrator is designed to perform extremely accurate orbital integrations of gravitational few-body systems \citep{Mikkola2006,Mikkola2008}. The equations of motion of the system are time-transformed by extending the phase space to include the original time parameter as a coordinate together with the corresponding conjugate momentum, equal to the binding energy of the system. A new independent variable is then introduced through a Poincar\'{e} time transformation. With a specific choice of the time transformation function \citep{Mikkola1999a,Preto1999}, the new Hamiltonian and the equations of motion are separable so that the system can be integrated using a leapfrog method. This surprisingly yields an exact orbit for the Keplerian two-body problem even for collision orbits. The only error is in the time coordinate, or the phase of the Keplerian binary. However, this error can be removed by a further modification of the Hamiltonian \citep{Mikkola2002a}, yielding an exact solver, up to machine precision.

We start with the standard N-body Hamiltonian $H$ defined as
\begin{equation}
    H = T - U = \sum_\mathrm{i} \frac{1}{2} m_\mathrm{i} \norm{\vect{v_\mathrm{i}}}^2 - \sum_\mathrm{i} \sum_\mathrm{j>i}  \frac{G m_\mathrm{i} m_\mathrm{j}}{\norm{ \vect{r}_\mathrm{j} - \vect{r}_\mathrm{i} } }
\end{equation}
in which $T$ is the kinetic energy and $U$ is the force function, equal to negative of the potential energy. This Hamiltonian yields the familiar Newtonian equations of motion for the particle positions $\vect{r}_\mathrm{i}$ and velocities $\vect{v}_\mathrm{i}$:
\begin{equation}
\begin{aligned}
    \derfrac{\vect{r}_\mathrm{i}}{t} &= \vect{v}_\mathrm{i}\\
    \derfrac{\vect{v}_\mathrm{i}}{t} &= \vect{a}_\mathrm{i} = G \sum_\mathrm{i \neq j} m_\mathrm{j} \frac{  \vect{r}_\mathrm{j} - \vect{r}_\mathrm{i}  }{ \norm{ \vect{r}_\mathrm{j} - \vect{r}_\mathrm{i} }^3 }
\end{aligned}
\end{equation}
in which we have introduced the Newtonian accelerations $\vect{a}_\mathrm{i}$. Possible additional acceleration terms, such as external perturbations $\vect{f}_\mathrm{i}$ depending only on particle positions, or velocity-dependent perturbations $\vect{g}_\mathrm{i}(\vect{v})$, such as post-Newtonian corrections, can be directly added to the Newtonian accelerations, yielding $\vect{a}_\text{i}\rightarrow\vect{a}_\mathrm{i} + \vect{f}_\mathrm{i} + \vect{g}_\mathrm{i}(\vect{v})$.

Next, we perform the time transformation \citep{Mikkola1999a,Preto1999}. A fictitious time $s$ is introduced as a new independent variable. The original independent variable, physical time $t$, is promoted to a coordinate of the phase space of the system, while the binding energy of the system $B = -H$ becomes the corresponding conjugate momentum. The old and new time variables are related by the infinitesimal time transformation
\begin{equation}\label{eq: ttl}
    \derfrac{t}{s} = \frac{1}{\alpha U + \beta \Omega + \gamma}
\end{equation}
in which the parameter triplet $(\alpha,\beta,\gamma)$ determines the type of regularisation. $\Omega$ is an arbitrary real-valued function of coordinates, such as the force function for the least massive particles in the system \citep{Mikkola2002b}. With the triplet $(1,0,0)$ the method becomes the logarithmic Hamiltonian (LogH) method of \citet{Mikkola1999a} and \citet{Preto1999}, while $(0,1,0)$ corresponds to the time-transformed leapfrog (TTL) introduced in \citet{Mikkola2002b}. The ordinary non-regularised leapfrog is obtained by choosing the triplet $(0,0,1)$. Of all the possible choices, \cite{Mikkola2006} recommend using the logarithmic Hamiltonian option $(1,0,0)$ for its superior numerical performance.

Using the logarithmic Hamiltonian time transformation, the equations of motion for the system become
\begin{equation}
\begin{aligned}
\derfrac{t}{s} &= \frac{1}{T+B}\\
\derfrac{\vect{r}_\mathrm{i}}{s} &= \frac{1}{T+B} \vect{v}_\mathrm{i}
\end{aligned}
\end{equation}
for the coordinates and
\begin{equation}
\begin{aligned}
\derfrac{\vect{v}_\mathrm{i}}{s} &= \frac{1}{U} (\vect{a}_\mathrm{i} + \vect{f}_\mathrm{i} + \vect{g}_\mathrm{i} (\vect{v}) )\\
\derfrac{B}{s} &= - \frac{1}{U} \sum_\mathrm{i} m_\mathrm{i} \vect{v}_\mathrm{i} \cdot (\vect{f}_\mathrm{i} + \vect{g}_\mathrm{i}(\vect{v}))
\end{aligned}
\end{equation}
for the velocities. In this discussion we omit the $\Omega$ function and its velocity conjugate. For an unperturbed Newtonian system i.e. $\vect{f}_\mathrm{i} = \vect{g}_\mathrm{i}(\vect{v}) = 0$ the derivatives of the coordinates depend only on the velocities and vice versa, thus a leapfrog algorithm can be constructed in a straightforward manner. With non-zero external tidal perturbations $\vect{f}_\mathrm{i}$ the derivative of the binding energy $B$ depends on the particle velocities, but the dependence is only linear and can thus be analytically integrated over the time-step (see e.g. the Appendix A and B of \citealt{Rantala2017}).

An explicit leapfrog algorithm cannot be constructed if the post-Newtonian accelerations $\vect{g}_\mathrm{i}(\vect{v})$ are non-zero. One can in practise approach the problem by iterating the implicit equations of motion, but this is very inefficient. An efficient post-Newtonian leapfrog algorithm with velocity-dependent accelerations can be implemented by extending the phase space of the system with auxiliary velocities $\vect{w}_\mathrm{i}$. A single leapfrog velocity update (kick) is replaced by an alternating combination of auxiliary and physical kicks, performed in a standard leapfrog manner. For additional details of the auxiliary velocity procedure see \cite{Hellstrom2010} and its generalisation by \cite{Pihajoki2015}.

Nevertheless, the LogH integrator on its own is not accurate enough for high-precision solutions of general N-body systems, even though the systems are regularised against collision singularities. The LogH leapfrog must be supplemented with additional numerical techniques such as chained coordinates and extrapolation techniques. These two methods are introduced in Section \ref{Section: chainedcoord} and Section \ref{Section: gbsold}, respectively.

\subsection{Chained coordinate system}\label{Section: chainedcoord}

In AR-CHAIN the chained inter-particle coordinate system does not play a role in the regularisation procedure itself, unlike in the earlier KS-CHAIN regularisation. However, numerical experiments \citep{Mikkola1999a,Mikkola1999b} have shown that the chained coordinate system is very useful in increasing the numerical accuracy of the method by significantly reducing the numerical floating point error.

When constructing the chained coordinates one first finds the shortest inter-particle coordinate vector of the N-body system. These two particles become the initial tail and head of the chain. Next, the particle closest to either the tail or the head of the chain is found among the non-chained particles. This particle is added as the new tail or head of the chain, depending which is end of the chain is closer. The process is repeated until all particles are in the chain.

Labelling the particles starting from the tail of the chain, the inter-particle position, velocity and various acceleration vectors become
\begin{equation}\label{eq: chained}
\begin{aligned}
    \vect{X}_\mathrm{k} &= \vect{r}_\mathrm{j_k} - \vect{r}_\mathrm{i_k} \equiv \vect{r}_\mathrm{k+1} - \vect{r}_\mathrm{k}\\
    \vect{V}_\mathrm{k} &= \vect{v}_\mathrm{j_k} - \vect{v}_\mathrm{i_k} \equiv \vect{v}_\mathrm{k+1} - \vect{v}_\mathrm{k}\\
    \vect{A}_\mathrm{k} &= \vect{a}_\mathrm{j_k} - \vect{a}_\mathrm{i_k} \equiv \vect{a}_\mathrm{k+1} - \vect{a}_\mathrm{k}\\
    \vect{F}_\mathrm{k} &= \vect{f}_\mathrm{j_k} - \vect{f}_\mathrm{i_k} \equiv \vect{f}_\mathrm{k+1} - \vect{f}_\mathrm{k}\\
    \vect{G}_\mathrm{k} &= \vect{g}_\mathrm{j_k} - \vect{g}_\mathrm{i_k} \equiv \vect{g}_\mathrm{k+1} - \vect{g}_\mathrm{k}
\end{aligned}
\end{equation}
in which the last expression on the right-hand side describes the relabelling of the particle indexes along the chain. Note that there are $N_\mathrm{part}-1$ inter-particle vectors for a system of $N_\mathrm{part}$ bodies. The equations of motion for the chained coordinates then become 
\begin{equation}
\begin{aligned}
\derfrac{t}{s} &= \frac{1}{T+B}\\
\derfrac{\vect{X}_\mathrm{i}}{s} &= \frac{1}{T+B} \vect{V}_\mathrm{i}
\end{aligned}
\end{equation}
while the velocity equations can be expressed as 
\begin{equation}
\begin{aligned}
\derfrac{\vect{V}_\mathrm{i}}{s} &= \frac{1}{U} (\vect{A}_\mathrm{i} + \vect{F}_\mathrm{i} + \vect{G}_\mathrm{i})\\
\derfrac{B}{s} &= - \frac{1}{U} \sum_\mathrm{i} m_\mathrm{i} \vect{v}_\mathrm{i} \cdot (\vect{f}_\mathrm{i} + \vect{g}_\mathrm{i}).
\end{aligned}
\end{equation}
It is worthwhile to note that the derivative of the binding energy $B$ is in fact easier to evaluate by using the original coordinate system than the chained one. For this the chained velocities need to be transformed back into the original coordinate system during the integration. 

Finally, the chained coordinate vectors are needed to evaluate the accelerations $\vect{a}_\mathrm{i}$ and $\vect{g}_\mathrm{i}(\vect{v})$. We use the chained coordinates for computing the separation vectors for $N_\mathrm{d}$ closest particles in the chain structure while the original vectors are used for more distant particles, i.e.
\begin{equation}\label{eq: close-chain}
    \vect{r}_\mathrm{j} - \vect{r}_\mathrm{i} =
    \left\{\begin{aligned}
            &\vect{r}_j - \vect{r}_i & \text{if $\abs{i-j} > N_{\mathrm{d}}$} \\
            &\sum_{k=\min\{i,j\}}^{\max\{i,j\}-1}\hspace{-3ex}
\sgn(i-j)\vect{X}_k & \text{if $\abs{i-j} \leq N_{\mathrm{d}}$}.
\end{aligned}\right.
\end{equation}
Typically $N_\mathrm{d} = 2$ in the literature \citep{Mikkola2008}. In general, selecting values $N_\mathrm{d}>2$ for the separation parameter undermines the usefulness of the chain construct as the floating-point error begins to accumulate when summing many inter-particle vectors in Eq. \eqref{eq: close-chain}.

\subsection{GBS extrapolation method}\label{Section: gbsold}

Even though the chained LogH leapfrog with the time-transformed equations of motion yields regular particle orbits for all non-pathological initial conditions, the numerical integration accuracy is usually too low for high-precision applications. Thus, the chained leapfrog integrator must be accompanied by an extrapolation method to reach a high numerical accuracy \citep{Mikkola1999b}. A widely used method is the Gragg--Bulirsch--Stoer \citep{Gragg1965,Bulirsch1966} or GBS extrapolation algorithm. The GBS extrapolation method can only be used with integrators that have an error scaling containing only even powers of the time-step, but fortunately many simple low-order integrators such as the mid-point method and the chained leapfrog fulfil this requirement. In this study we use only leapfrog-type integrators with the GBS algorithm.

In general, when numerically solving initial value problems for differential equations the numerical solution will converge towards the exact solution when the step-size $h$ of the numerical method is decreased. The numerical accuracy of integrators with an extrapolation method is based on this fact. The key idea of the GBS extrapolation is to successively integrate the differential equation over an interval $H$ using an increasing number of substeps $n$. The integrations are carried out in small steps of length $h=H/n$ using a suitable numerical method, and the results are then extrapolated to $h \rightarrow 0$. Different substep division sequences $n_\mathrm{k}$ have been studied in the literature to achieve converged extrapolation results with a minimum computational cost \citep{Press2007}. Popular options include the original GBS sequence \citep{Bulirsch1966} defined as
\begin{equation}\label{eq: sec-gbs}
    \{n_\mathrm{k}\} = \{2,4,6,8,12,16,24,32,48,64,96\dots\}
\end{equation}
i.e. $n_\mathrm{k} = 2 n_\mathrm{k-2}, k>2$ and the so-called Deuflhard sequence \citep{Deuflhard1983} of even numbers
\begin{equation}\label{eq: seq-deuflhard}
    \{n_\mathrm{k}\} = \{2,4,6,8,10,12,14,16,18,20,22,\dots\}
\end{equation}
with $n_\mathrm{k} = 2k$. In our \mstar{} code and throughout this paper we use the Deuflhard sequence.

The chained leapfrog sequence with $n \geq 1$ substeps can be written as
\begin{equation}
    \vect{D}\left( \frac{h}{2n} \right) \left[ 
    \vect{K}\left( \frac{h}{n} \right) \vect{D}\left( \frac{h}{n} \right)      \right]^{n-1} \vect{K}\left( \frac{h}{n} \right) \vect{D}\left( \frac{h}{2 n} \right),
\end{equation}
where the drift $\vect{D}(h)$ operators advance the coordinates and the kick operators $\vect{K}(h)$ the velocity-like variables by the time-step $h$. The GBS algorithm starts by computing the first few substep divisions $n_\mathrm{k}$ with the leapfrog after which a rational function or a polynomial extrapolation of the integration results to $h \rightarrow 0$ is attempted. The error control is enabled with the convergence criterion
\begin{equation}\label{eq: gbs-criterion}
    \frac{\norm{\Delta S_k}}{\norm{S(s) + \frac{h}{n}S'(s)}} \leq \etabs,
\end{equation}
where $S$ is any dynamical variable of the system, $\Delta S_k$ is the extrapolation error estimate after the $k$'th substep sequence and $S(s)$ and $S'(s)$ are the value of the dynamical variable and its time derivative obtained after the last complete time-step $H$. The GBS tolerance $\etabs$ is a user-given free input parameter. The extrapolation can typically be carried out successfully even if $\etabs$ is set near the double-precision floating-point precision $\etabs \sim 10^{-16}$. In the literature, the most typical values of the accuracy parameter are set in the range of $10^{-12} \leq \etabs \leq 10^{-6}$

If convergence is not reached after the first few leapfrog step divisions, the GBS algorithm proceeds to the next substep division and tries the extrapolation again until convergence, or until the maximum number of divisions $n_\mathrm{max}=n_{k_\text{max}}$, is reached. In the case of no convergence after the $k_\mathrm{max}$'th substep sequence, the original step $H$ is halved and the process is started again from the beginning. After convergence, the criterion for the next time-step $H_{i+1}$ after convergence is \citep{Press2007}
\begin{equation}\label{eq: hnext}
H_\mathrm{i+1} = a_\mathrm{GBS} \left( \frac{\etabs}{\epsilon_\mathrm{k}}\right)^{1/(2k-1)} H_\mathrm{i},
\end{equation}
where $\epsilon_\mathrm{i}$ is the maximum error in the dependent variables from the previous step, $a_\text{GBS}\in(0,1]$ is a safety factor \citep{Hairer2008}, and $k$ is the substep sequence at which convergence was achieved. The GBS algorithm also monitors whether trying convergence at different $k$, or equivalently, changing the order $2k-1$ of the method, would lead to convergence with a smaller work-load. The time-step $H$ is then adjusted accordingly, to bring the $k$ where convergence is expected to the optimal value \citep{Press2007}.

Finally, it should be noted that the extrapolation algorithm used in \archain{} described above is serial in nature, even though the particle accelerations in the individual leapfrog sequences can be computed in parallel as in \citep{Rantala2017}. Thus, computing a large number of subsequent subdivision counts or reaching the maximum subdivision count $k_\mathrm{max}$ without convergence and restarting with a smaller time-step $H/2$ can be very time-consuming.

\subsection{Iteration to exact physical time}\label{Section: iter}

We next describe an improvement of the time iteration procedure in the new \mstar{} integrator over our older \archain{} version. Consider the integration of the equations of motion of an N-body system over a time interval $\Delta t$. With the standard leapfrog there is no problem in arriving at the correct end time, but with the time transformed leapfrog one has to be more careful (e.g. \citealt{Mikkola1997}). 

Integrating the time transformation of Eq. \eqref{eq: ttl} over a time interval $H=\Delta t$ with the parameter triplet $(1,0,0)$ yields
\begin{equation}\label{eq: delta-s}
    \Delta s = \int_0^{\mathrm{\Delta t}} U dt = G \sum_{\mathrm{i}} \sum_{\mathrm{j > i}} m_\mathrm{i} m_\mathrm{j} \int_0^{ \mathrm{\Delta t}} \frac{dt}{\norm{ \vect{r}_\mathrm{j} - \vect{r}_\mathrm{i} }}.
\end{equation}
One can in principle approach these $N_\mathrm{part}(N_\mathrm{part}-1)/2$ integrals in the formula by using the Stumpff-Weiss method which assumes that all the particle motions during the time interval are Keplerian \citep{Stumpff1968}. However, a simple approximation
\begin{equation}\label{eq: t-aika}
    \Delta s = U \Delta t
\end{equation}
typically provides sufficiently accurate results, especially when the time interval $\Delta t$ is short. 

In our \archain{} implementation \citep{Rantala2017} we begin the regularised integration by estimating the amount of fictitious time $\Delta s$ based on the smallest Keplerian time-scale $P_\mathrm{Kepler}$ of the system. Using Eq. \eqref{eq: t-aika} we have $\Delta s = q U P_\mathrm{Kepler}$ in which $q$ is a safety factor $0<q\leq 1$. After the first step we let the GBS algorithm decide the step size until we exceed the output time $\Delta t$. Then, we take additional GBS iteration steps towards the correct exact time until the relative difference between the integrated time and $\Delta t$ is small enough, typically $\sim 10^{-4}$--$10^{-6}$. This requires $2$ to $5$ iteration steps in most cases.

We improve the time iteration procedure for our new integrator as the GBS steps are expensive and one should try to converge to the correct physical time with as few iterations as possible. For the first step we already use Eq. \eqref{eq: t-aika} multiplied by a safety factor instead of the Keplerian criterion. Next we assume that after a GBS step has been taken we have arrived at the physical time $0<\tau<\Delta t$. As the GBS routine suggests a new time step $\Delta s$ we check that the estimated time coordinate after the next step $\tau + \Delta s / U$ does not exceed $\Delta t$. If it does, we set the next fictitious time step to $\Delta s = U(\Delta t - \tau)$. If we still end up to time $\tau>\Delta t$ after integration we proceed as in the old method. This procedure typically requires a few expensive iteration steps less than the approach we used in our previous \archain{} implementation. The speed-up gained by the updated iteration procedure depends on $\Delta t$ and the N-body system, and the maximum expected speed-up occurs in the integration of very short time intervals.

\section{Minimum spanning tree coordinates} \label{Section: MST}
\subsection{Shortest Hamiltonian paths and minimum spanning trees}\label{Section: graph}

The properties of chained inter-particle coordinate systems can be conveniently expressed by using the language of graph theory (e.g. \citealt{Harary1969}). A graph $G = (V,E)$ is an ordered pair of vertices $V$ and edges $E$. An edge is determined by the vertices it connects, i.e. $E_{ij} = (V_\mathrm{i},\:V_\mathrm{j})$. For our purposes the N-body particles are the vertices of the graph while the inter-particle vectors correspond to the edges of the graph. A graph with $N$ vertices and all possible $N (N-1)/2$ edges is called complete. Complete graphs are also necessarily connected as any vertex can be reached from any other vertex of the graph. Each edge $E$ is weighted with a non-negative real number $w$. We set the weights of the edges by calculating the Euclidean norm i.e. the length of the inter-particle vector corresponding to each edge.

In graph theory a path is a sequence of distinct vertices. A Hamiltonian path is a path which visits each vertex of the graph exactly once. In a complete graph Hamiltonian paths are guaranteed to exist. We note here that the chain structures of \archain{} are in fact Hamiltonian paths. The problem of finding whether a Hamiltonian path exists in a given graph is NP-complete i.e. in the practical point of view meaning no solution in polynomial time $\bigO(N^k)$ with $k>0$ exists. Furthermore, it can be shown that there is no polynomial time algorithm to find the shortest Hamiltonian path of a complete graph (i.e. shortest chain) either. The computational upper limit of a brute force approach to constructing the shortest chain scales as $\bigO(N!)$. Thus, strictly speaking our procedure in finding the chain of inter-particle vectors in Section \ref{Section: chainedcoord} corresponds to finding only the approximately shortest Hamiltonian path of the system. Consequently, there are usually more optimal chained coordinate systems than the one we find with our chain construction algorithm but going through all the possible chain configurations is not feasible.

A spanning tree $T = (V,E_\mathrm{T})$ of the graph $G$ is a subgraph of $G$ connecting all the vertices of the original graph with a minimum possible number of its edges. A spanning tree connecting $N$ vertices has $N-1$ edges, the same as the number of inter-particle vectors in the chain structure of \archain. In addition, $T$ is a minimum spanning tree (\MST{}) of $G$ if the sum of the edge weights $w$ in $T$ is the smallest among all the possible spanning trees of $G$. It turns out that if the graph $G$ has unique edge weights $w$ there is a unique minimum spanning tree $T$. This is usually the case in practical applications of minimum spanning trees, such as our N-body systems. Unlike for the shortest Hamiltonian path problem, there are algorithms which find the \MST{} for a given complete graph in a polynomial time.

There are two important graph properties which aid in finding the \MST{} or in filtering out edges which are certainly not in the \MST{} of a graph. The first is the cycle property. A cycle $C$ of graph $G$ is a path which forms a loop. The cycle property states that the edge $E_\mathrm{i}$ with the highest weight in any cycle $C$ in $G$ cannot be in the \MST{} of $G$. The second property is the cut property. Divide the vertices $V_\mathrm{i}$ of $G$ arbitrarily into two distinct groups. The essence of cut property is that the edge $E_\mathrm{ij}$ with the minimum weight connecting the two vertex groups is necessarily in the \MST{} of $G$.

\subsection{Finding the MST: Prim's algorithm}

The problem of efficiently finding the minimum spanning tree of a given graph has been extensively studied in the literature, the classic solutions to the problem being found by \cite{Boruvka1926}, \cite{Kruskal1956} and \cite{Prim1957}. We select the classic Prim's algorithm due to its relative simplicity and the fact that the algorithm somewhat resembles the chain construction in our \archain{} implementation with the difference that the chain is now allowed to branch. In addition, Prim's algorithm makes labelling the edge vectors easier for our purposes than the other two classic algorithms. 

Prim's algorithm proceeds as follows. First, one selects a single vertex $V_\mathrm{i}$ of the graph $G$. For N-body systems we suggest starting from the particle spatially closest to the centre-of-mass of the system. Next, the edge $E_\mathrm{ij}$ with a minimum weight $w$ connected to the first vertex is found and added as the first edge to the \MST{}. Then the algorithm proceeds by finding the successive minimum weight edges among the edges in $G$ connected to the \MST{} and adds them into the \MST{} until all vertices are connected with the \MST{}. Our parallel implementation uses sorted adjacency lists on different tasks to effectively find the consecutive edges to add to the \MST{}. For complete graphs even the most sophisticated \MST{} finding algorithms typically scale as $\bigO(N^2)$ as the Prim's algorithm does.

\begin{figure}
\includegraphics[width=\linewidth]{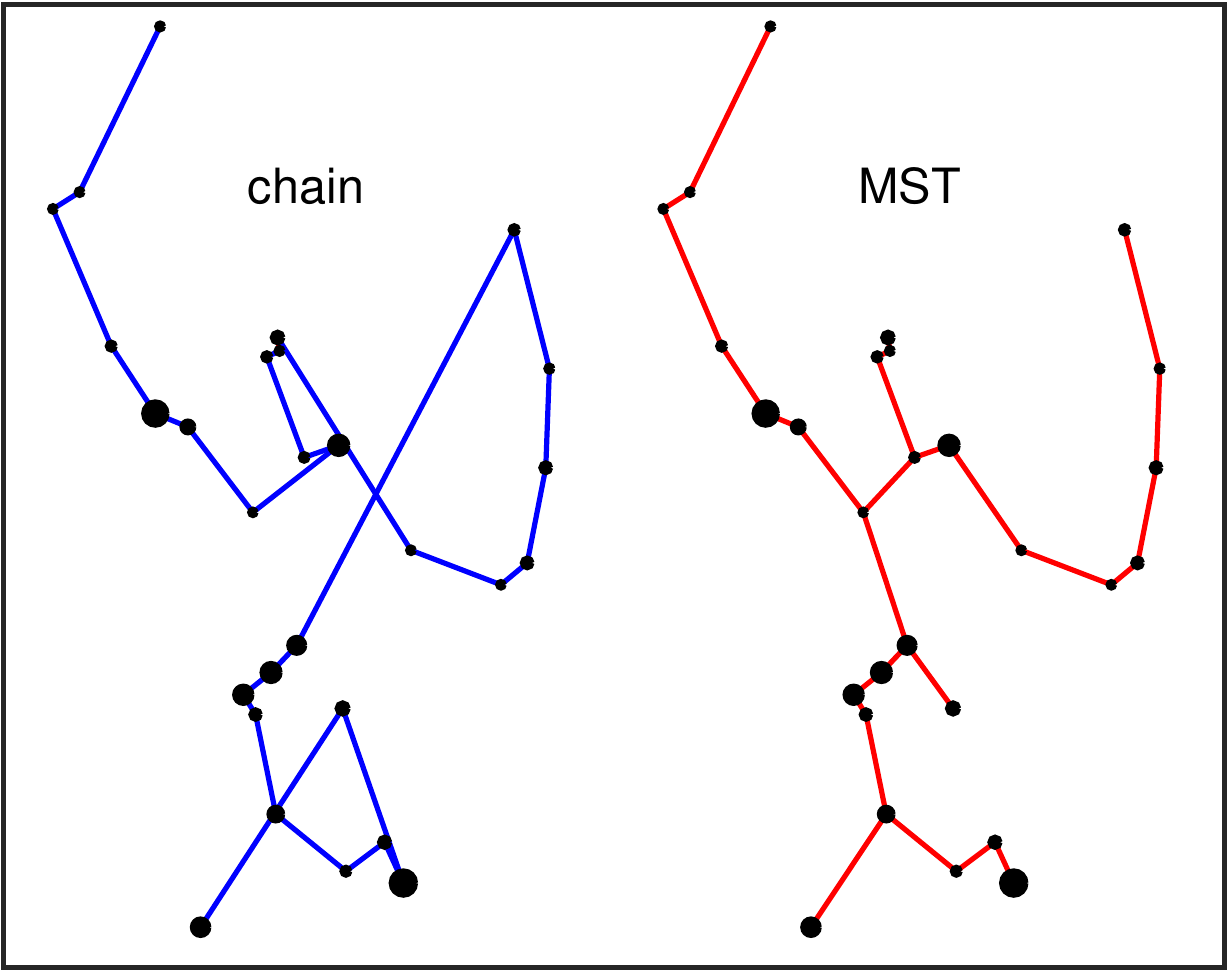}
\caption{A two-dimensional illustration highlighting the difference between a chained coordinate system and the minimum spanning tree coordinates. Both the chain and the \MST{} are constructed on a collection of $28$ points corresponding to the locations of the brightest stars in the familiar constellation of Orion. The total length of the \MST{} edges is smaller than the length of the chain. The chain also occasionally suffers from the fact that spatially close particles might be distant in the chain, which the \MST{} avoids by branching its tree structure.}
\label{fig1}
\end{figure}

The crucial difference between the chain construction and Prim's algorithm manifests itself here. In the chain construction it is allowed to add new inter-particle vectors only to the tail and the head of the chain while in Prim's algorithm it is allowed to add new edges to any location in the \MST{}. This ensures that spatially close N-body particles are always located near each other in the \MST{} data structure, which is necessarily not the case in the chain. The differences between a chain and a \MST{} built on a same group of particles is illustrated in Fig. \ref{fig1}.

\subsection{Finding the MST: divide-and-conquer method}

It is possible to quickly find spanning trees $T$ with a total weight very close to the total weight of the \MST{} of the graph (e.g \citealt{Wang2009, Zhong2013}). For simplicity we also refer to these approximate minimum spanning trees as \MST{}s  throughout this study.

Our preferred method is the divide-and-conquer approach to the Prim's algorithm. First one divides the original graph $G$ into $\sim \sqrt{N}$ subgraphs $G'$. We use a simple octree-based spatial partition to find these subgraphs. A so-called meta-graph $G''$ is formed as the contracted subgraphs as its vertices and including all possible edges between the vertices. The edge weights of the meta-graph are the minimum edge weights between the subgraphs by the cut property. Next we use Prim's algorithm to construct the minimum spanning trees of each $G'$ and the meta-graph $G''$. To speed up the local \MST{} construction we eliminate edges which cannot be in the final \MST{} using the cycle property before applying Prim's algorithm. Now we have all the edges of the total \MST{} which are then ordered and labelled by performing a standard tree walk.

We find that the spanning trees $T$ found using the approximate algorithm are typically $5\%$ longer than the true \MST{} of the graph. However, this difference is not a serious problem as the spanning trees are locally true \MST{}s of the subgraphs. Furthermore, our motivation to use \MST{}s is to minimise numerical error originating mostly from computation operations involving spatially close particles by choosing a clever local coordinate system. In addition, the divide-and-conquer approach is faster than the original Prim's algorithm. Concerning the speed of the algorithms we find that in our code it is always profitable to use the divide-and-conquer method when the particle number is $N\gtrsim10^3$. With smaller particle numbers both approaches yield very similar results as the wall-clock time elapsed in the \MST{} construction is negligible.

\subsection{MST as a coordinate system}\label{Section: MST-coord}

As both the chain and the \MST{} consist of $N_\mathrm{part}-1$ inter-particle vectors, the \MST{} coordinate system can be constructed with a recipe similar to the chained coordinates in Eq. \eqref{eq: chained} with relatively small modifications to the procedure.

\begin{figure}
\includegraphics[width=\linewidth]{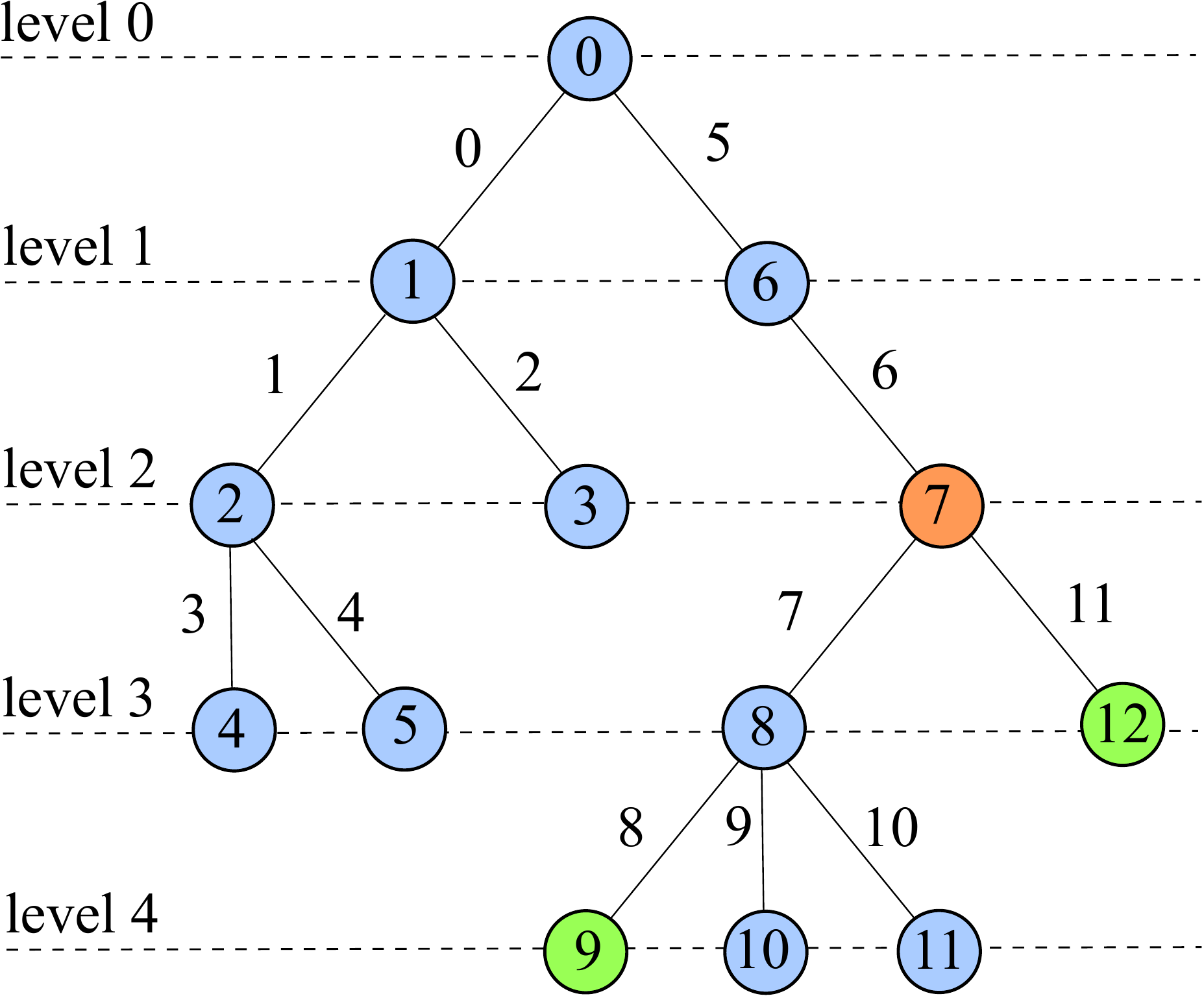}
\caption{A tree structure with $12$ vertices, $11$ edges and $5$ levels. The parent vertex of vertices $V_\mathrm{2}$ and $V_\mathrm{3}$ is the vertex $V_\mathrm{1}$. In addition, the vertex $V_\mathrm{1}$ is the 2nd ancestor of vertex $V_\mathrm{4}$. The lowest common ancestor of vertices $V_\mathrm{9}$ and $V_\mathrm{12}$ (both in green) is the vertex $V_\mathrm{7}$ (in orange).}
\label{fig2}
\end{figure}

First, we need to index the \MST{} edge vectors $E_\mathrm{ij}$ as in Eq. \eqref{eq: chained}. In the chain construct setting the chained indices is straightforward as one simply starts the indexing from the tail of the chain and proceeds towards the head. In the \MST{} indexing is more complicated because of the branching of the \MST{} structure. However, we can take full advantage of the fact that the \MST{} is a tree structure. The first chosen vertex $V_\mathrm{0}$ is the root vertex of the \MST{} and gets the index and level zero in the \MST{} i.e. $L(V_\mathrm{0}) = 0$. We index the subsequent vertices and edge vectors in the order they were added to the \MST{}. The vertex levels and the parents of the vertices are assigned simultaneously as well. A simplified illustration describing the indexing of our \MST{} structure is presented in Fig. \ref{fig2}. After the indexing the inter-particle vectors corresponding to the \MST{} edges can be computed just as in Eq. \eqref{eq: chained} with the additional rule that all the inter-particle vectors point away from the root vertex, just as the chained vectors are oriented away from the tail towards the head of the chain.

Next, we generalise Eq. \eqref{eq: close-chain} to determine the rules when two vertices are close to each other in the \MST{}, i.e. within $N_\mathrm{d}$ edges of each other, just as in the chain construct. For the force calculation of the nearby vertices the \MST{} edge vectors are used while the original coordinate system is used for the rest of the vertex pairs. The criterion of two vertices $V_\mathrm{i}$ and $V_\mathrm{j}$ being within $N_\mathrm{d}$ \MST{} edges of each other can be conveniently expressed by using the lowest common ancestor (LCA) of the two vertices. The parent vertex is the 1st ancestor of the vertex, the 2nd ancestor is the parent of the parent vertex, and so on. The LCA is the vertex among the common ancestors of both $V_\mathrm{i}$ and $V_\mathrm{j}$ which has the highest level in the \MST{}. We label this vertex $V_\mathrm{LCA}$. Note that $V_\mathrm{i}$ or $V_\mathrm{j}$ itself may be the $V_\mathrm{LCA}$ of the vertex pair. Now we can state that if
\begin{equation}\label{eq: LCA}
|L(V_\mathrm{i})-L(V_\mathrm{LCA})| + |L(V_\mathrm{j})-L(V_\mathrm{LCA})| \leq N_\mathrm{d}
\end{equation}
the two vertices $V_\mathrm{i}$ and $V_\mathrm{j}$ are close to each other in the \MST{}. Here $L(V_\mathrm{i})$ again signifies the level of the vertex $V_\mathrm{i}$. 

Finally, we write down the recipe for selecting which vectors to use in the force calculation. If $V_\mathrm{j}$ and $V_\mathrm{i}$ are within $N_\mathrm{d}$ edges of each other, we simply walk the \MST{} from $V_\mathrm{i}$ to $V_\mathrm{j}$ via $V_\mathrm{LCA}$ and sum the traversed edge vectors to obtain the \MST{} separation vector $\vect{X}_\mathrm{k}$, just as in Eq. \eqref{eq: close-chain} with the chain. If the condition of Eq. \eqref{eq: LCA} does not hold, we use the original coordinate vectors to compute the separations $\vect{r}_\mathrm{j} - \vect{r}_\mathrm{i}$.

\subsection{Numerical error from consecutive coordinate transformations}\label{Section: rebuild}

\begin{figure}
\includegraphics[width=\linewidth]{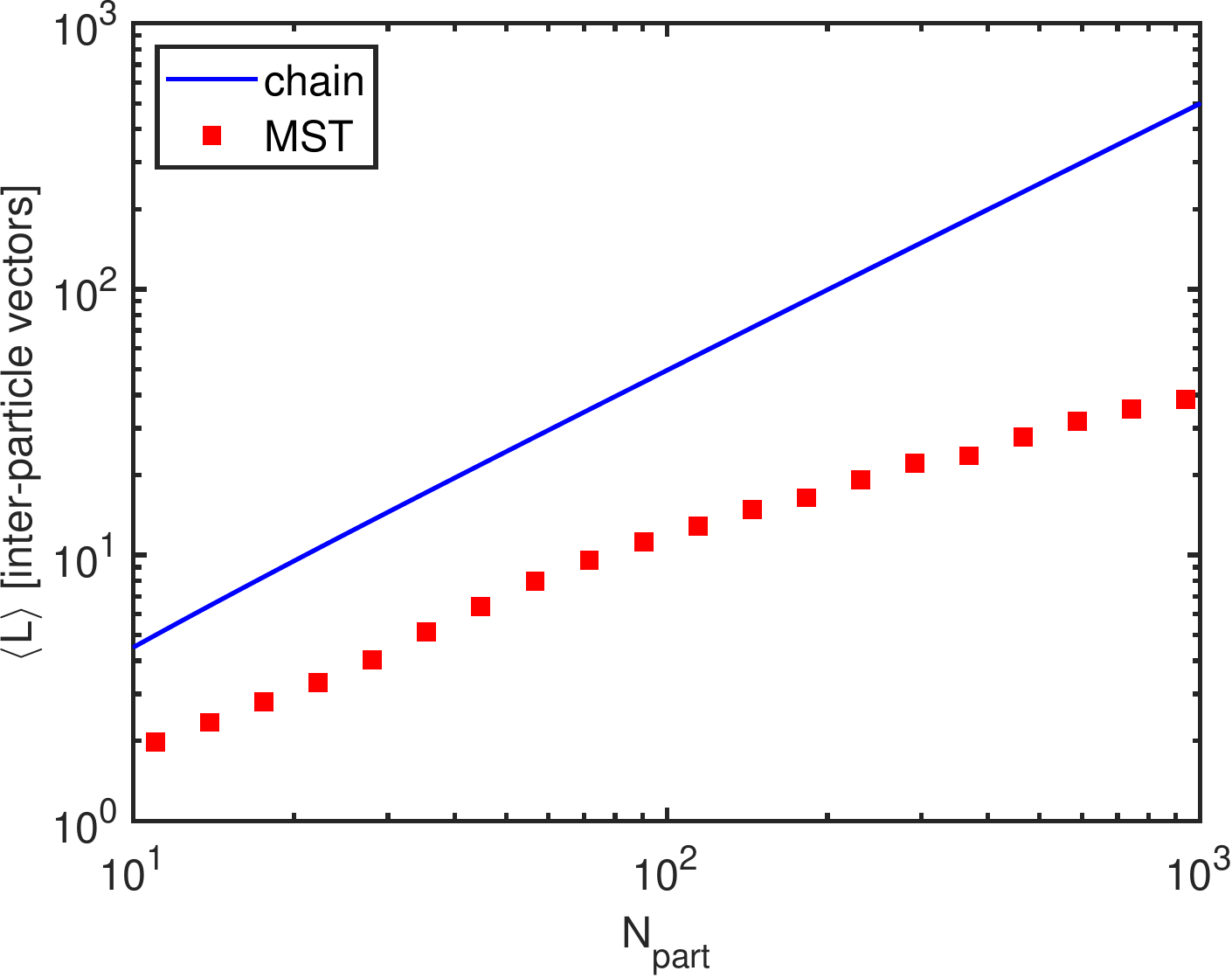}
\caption{Comparing the mean length $\langle L_\mathrm{chain} \rangle$ of the a chain structure (solid blue line) and the mean level $\langle L_\mathrm{MST} \rangle$ of particles in a \MST{} (red symbols). The particle distribution on which the coordinate systems are build follows a $\rho(r) \propto r^{-1}$ density profile. We see that the mean level of particles in the inter-particle coordinate structure is always lower in the \MST{} by a factor of $\sim 2$--$10$. Changing the underlying density profile has a very small effect on the result. The small mean particle level corresponds to a smaller level of floating-point error when constructing and deconstructing the coordinate systems as discussed in the text.}
\label{fig: mean_L}
\end{figure}

During integration the \MST{} (or chain) coordinate structure is frequently built, deconstructed and rebuilt to keep the coordinate structure up-to-date as the simulation particles move. For N-body systems with a large number of particles these coordinate transformations require a large number of summation operations (e.g. \citealt{Mikkola1993}), which introduces a new source of numerical error in the integration. This fact has received little attention in the literature thus far, most probably due to the fact that for few-body systems ($N_\mathrm{part}\lesssim10$) the accumulated summation errors always remain small. The chain construction and deconstruction both require summing on average $\langle L_\mathrm{chain} \rangle = (N_\mathrm{part}-1)/2$ coordinate vectors. For \MST{} the corresponding number is the mean level of the vertices in the \MST{}, i.e. $\langle L_\mathrm{MST} \rangle = \langle L(V_\mathrm{i}) \rangle$. Now the choice of the root vertex in Section \ref{Section: graph} becomes important. If the root vertex is spatially close to the centre-of-mass of the system, $\langle L_\mathrm{MST}\rangle$ should always be smaller than $\langle L_\mathrm{chain} \rangle$. We demonstrate this fact in Fig. \ref{fig: mean_L}. Our results show that $\langle L_\mathrm{chain}\rangle/\langle L_\mathrm{MST}\rangle \sim 2$--$10$ with particle numbers of $N_\mathrm{part} = 10$--$1000$. 

\begin{figure}
\includegraphics[width=\linewidth]{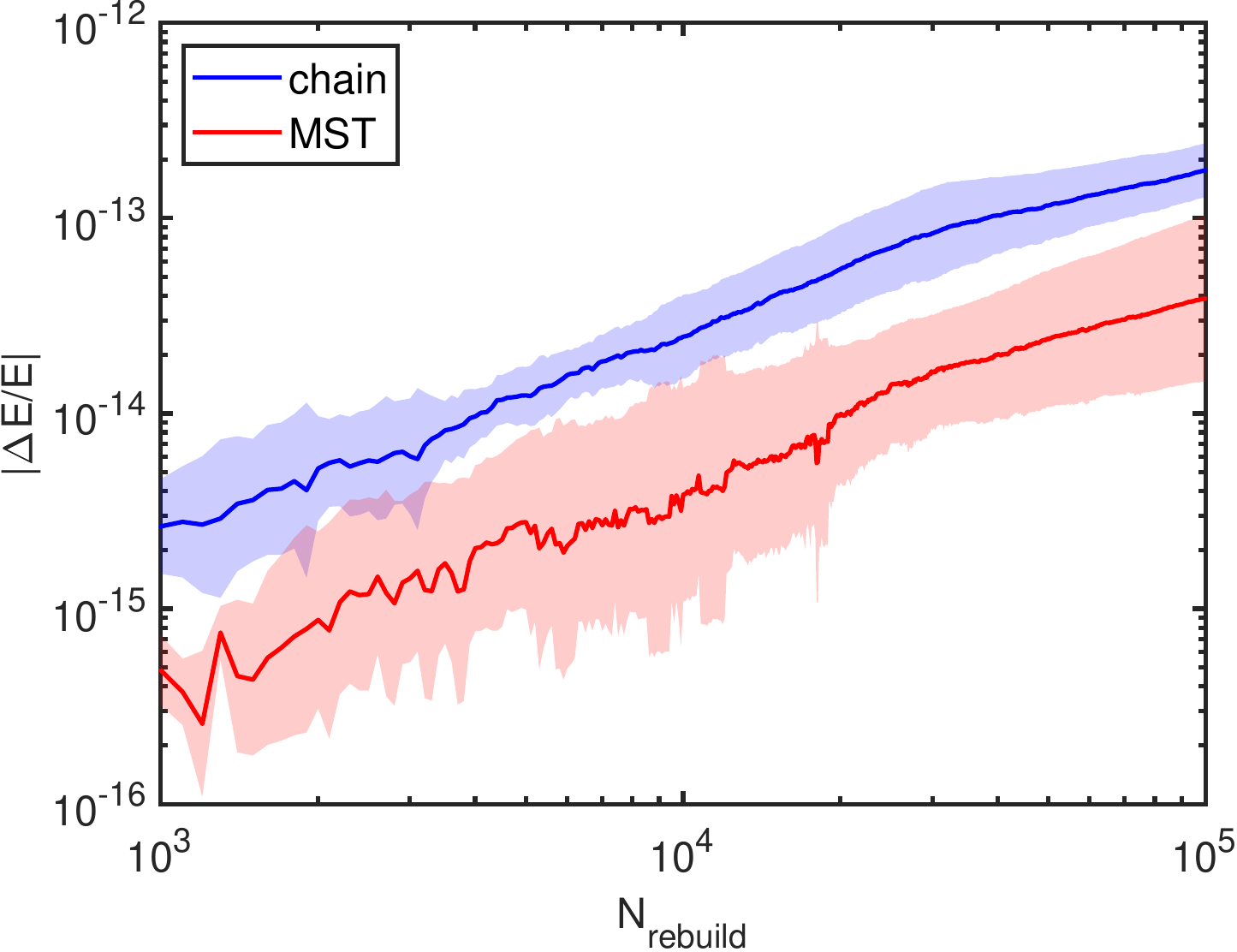}
\caption{The build, deconstruct and rebuild test of the two coordinate systems on a simple N-body system with $N_\mathrm{part}=379$ particles. With the chained coordinates (blue line) the energy error is approximately an order of magnitude larger than with the \MST{} coordinates (red line). The results are averages over $10$ N-body systems with differing random seeds. The coloured regions represent the scatter of one standard deviation.}
\label{fig: rebuild}
\end{figure}

This difference somewhat affects the numerical performance of the two algorithms. We perform an additional numerical test in which we build, deconstruct and rebuild the two coordinate systems consecutively $N_\mathrm{rebuild} = 10^5$ times while monitoring the accumulating numerical error. The N-body system on which the coordinate systems are built contains $N_\mathrm{part}=379$ particles with its full details presented in Section \ref{Section: ic}. The results of the rebuild test are presented in Fig. \ref{fig: rebuild}. The results indeed show that the \MST{} coordinate system accumulates less numerical error than the chained coordinates, the difference being approximately an order of magnitude in energy error. Apart from the difference of an order of magnitude the cumulative error behaves very similarly with the two coordinate systems. As the energy error in the chain coordinate test reaches $|\Delta E/E|\sim10^{-13}$ after $N_\mathrm{rebuild} = 10^5$ rebuilds we conclude that the \MST{} coordinate system is recommended for regularised N-body simulations requiring extremely high numerical accuracy.

Finally, we note that advanced floating-point summation methods such as the Kahan summation \citep{Kahan1965} or the Neumaier summation \citep{Neumaier1974} algorithm might be used to further diminish the numerical error from the coordinate transformation operations. However, the inclusion of such more advanced summation algorithms is left for future work, as our current \mstar{} code is numerically accurate enough for all current target applications.

\section{Parallel extrapolation method}\label{Section: parallel}
\subsection{Force loop parallelisation}\label{Section: para-force}

The most straightforward way to begin to parallelise a serial algorithmically regularised integrator is to parallelise the force computation of the code. The MPI parallelisation standard is adopted for this study. We use the basic parallelisation strategy in which the $\bigO(N_\mathrm{part}^2)$ iterations of the force computing loop are divided evenly for $N_\mathrm{force}$ MPI tasks, speeding up the loop calculation. However, the inter-task communication required to collect the final results after the force computations uses an increasing amount of CPU time when the number of MPI tasks is increased.

In a serial direct summation integrator using a GBS extrapolation method the total wall-clock time $T$ elapsed for the force calculation during the integration of a single step $H$ can be expressed as
\begin{equation}\label{eq: force-wallclock}
    T \approx \sum_\mathrm{k=1}^\mathrm{k_\mathrm{max}} n_\mathrm{k} t N_\mathrm{part}^2 = t N_\mathrm{part}^2  \sum_\mathrm{k=1}^\mathrm{k_\mathrm{max}} n_\mathrm{k},
\end{equation}
where $k_\mathrm{max}$ is the maximum number of GBS substep divisions and $t$ is a proportionality constant. Without the GBS algorithm the sum expression would not appear in the formula. Thus, the wall-clock time elapsed in the force calculation depends not only on the particle number $N_\mathrm{part}$ but also on the sequence $n_\mathrm{k}$ introduced in Eq. \eqref{eq: sec-gbs} and Eq. \eqref{eq: seq-deuflhard}.

We define the speed-up factor of the parallelised force computation as
\begin{equation}\label{eq: speedup-force}
S_\mathrm{force}(N_\mathrm{force}) = \frac{T_\mathrm{serial}}{T_\mathrm{parallel}(N_\mathrm{force})}
\end{equation}
in which $T_\mathrm{serial}$ and $T_\mathrm{parallel}$ are the respective wall-clock times elapsed during the force computation. Assuming ideal zero-latency communication between the tasks, the speed-up factor $S_\mathrm{force}$ of the parallelised force calculation scales initially linearly with the number of MPI tasks, i.e. $S_\mathrm{force} \propto N_\mathrm{force}$. The linear scaling behaviour continues until the number of tasks equals the particle number $N_\mathrm{force} = N_\mathrm{part}$ at which point one MPI task handles one particle. After this the speed-up factor $S_\mathrm{force}$ remains constant. With realistic non-instantaneous inter-task communication the flat scaling behaviour is reached with $N_\mathrm{force}$ well below the particle number $N_\mathrm{part}$ due to the increasing communication costs.

\begin{figure}
\includegraphics[width=\linewidth]{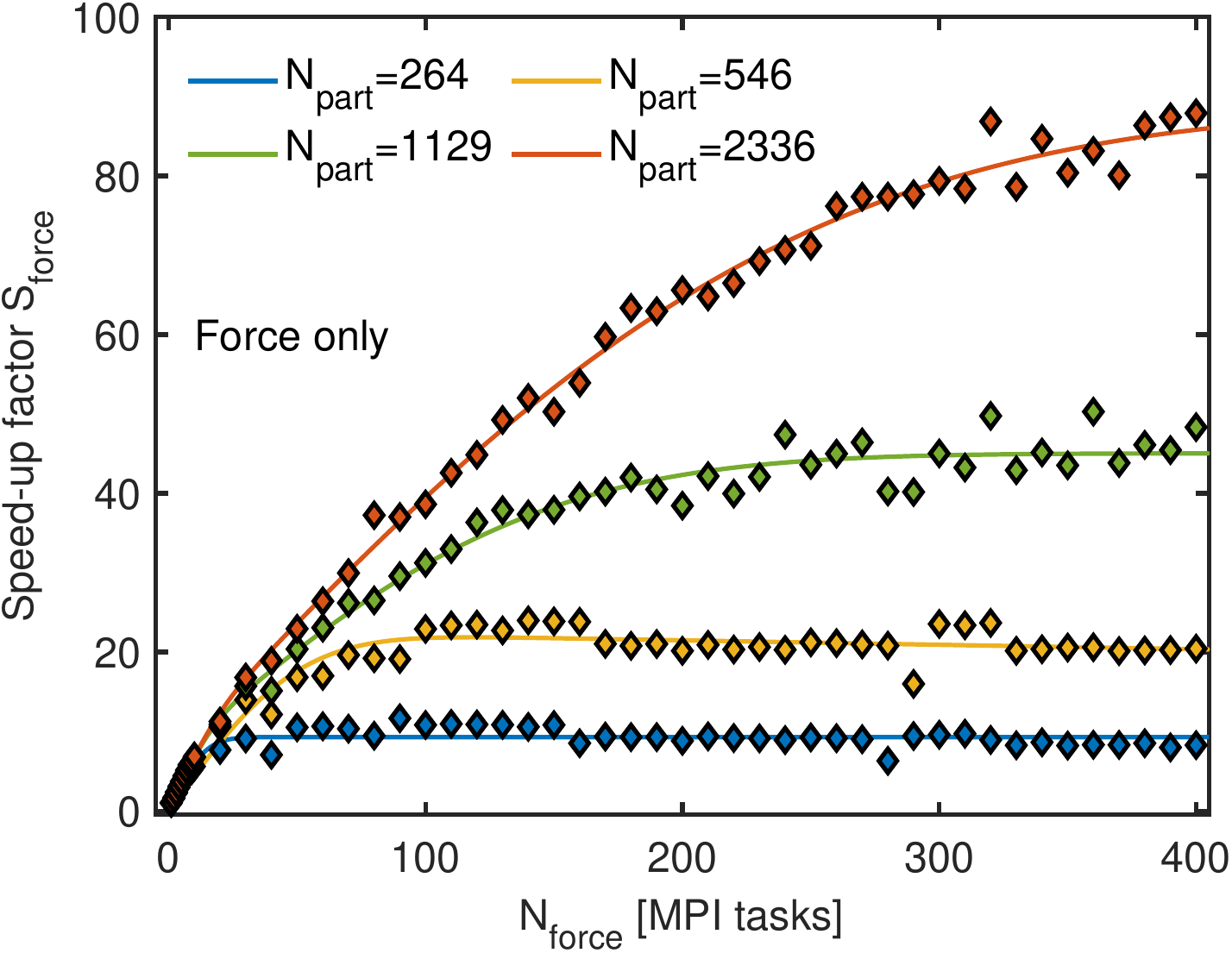}
\caption{The speed-up factor of the force computation $S_\mathrm{force}$ as a function of the number of MPI tasks $N_\mathrm{force}$ for four different N-body particle numbers $N_\mathrm{part}$. The symbols represent the measured force computation speed-up factors while the solid lines are error function fits (Eq. \ref{eq: errorfun}) to the speed-up data as described in the text. The speed-up factor initially grows linearly when $N_\mathrm{force} \ll N_\mathrm{part}$ and saturates to a roughly constant level after $N_\mathrm{force} \gtrsim 0.1\times N_\mathrm{part}$.}
\label{fig: theorspeedup-2}
\end{figure}

We illustrate the results of a series of force computation scaling tests in Fig. \ref{fig: theorspeedup-2}. The N-body systems in the test are selected from our sample of initial conditions with logarithmically spaced particle numbers in the range $10^1 \leq N_\mathrm{part} \leq10^4$. The results are averages over three different random realisations of the systems. In the scaling test we use $1 \leq N_\mathrm{force} \leq 400$ MPI tasks. 
With a small number of MPI tasks $(N_\mathrm{force} \lesssim 10)$ the speed-up factor increased linearly, as expected. After the linear phase $S_\mathrm{force}(N_\mathrm{force})$ flattens to a constant function at higher $N_\mathrm{force}$. Eventually the speed-up factor actually begins to decrease as the communication costs start to dominate over the time elapsed in the force computation. We define the maximum reasonable task number $N_\mathrm{force}^\mathrm{max}$ for N-body systems as the task number in which $\sim 95\%$ of the maximum speed-up has been achieved, i.e. $S_\mathrm{force}(N_\mathrm{force}^\mathrm{max}) = 0.95 \times S_\mathrm{force}^\mathrm{max}$. When $N_\mathrm{force} \gtrsim N_\mathrm{force}^\mathrm{max}$ the addition of subsequent MPI tasks has only a negligible effect on the speed-up of the force computation. We find in our scaling experiments that $N_\mathrm{force}^\mathrm{max}$ can be approximated with a simple relation
\begin{equation}
    N_\mathrm{force}^\mathrm{max} \approx q\times N_\mathrm{part}
\end{equation}
in which the constant factor is between $0.05 \leq q \leq 0.1$. In addition, the maximum force computation speed-up factor $S_\mathrm{force}^\mathrm{max}$ can be approximated with the formula
\begin{equation}
    \log_\mathrm{10}(S_\mathrm{force}^\mathrm{max}) \approx a_\mathrm{1} \log_\mathrm{10}(N_\mathrm{part})  - a_\mathrm{2}.
\end{equation}
with $a_\mathrm{1} \approx 0.505$ and $a_\mathrm{2} \approx -0.58$.
For quantifying the behaviour of the speed-up factor $S_\mathrm{force}$ in the 
intermediate range of MPI tasks between the linearly increasing and the constant speed-up factor a suitable fitting function is required. A suitable choice is the error function $\erf{}$ which has the correct asymptotic behaviour both with small and large values of its argument. We use fitting functions of the form 
\begin{equation}\label{eq: errorfun}
S_\mathrm{force}(N_\mathrm{force}) = \sum_\mathrm{i}^\mathrm{N_\mathrm{coeff}} b_\mathrm{i}\erf{(c_\mathrm{i} N_\mathrm{force})}    
\end{equation}
in which $b_\mathrm{i}$ and $c_\mathrm{i}$ are constant coefficients. As expected, $\sum_\mathrm{i}^\mathrm{N_\mathrm{coeff}} b_\mathrm{i} \approx S_\mathrm{force}^\mathrm{max}$ by definition. We find that using $N_\mathrm{coeff} \approx 1$--$2$ terms yields good results. The fit coefficients are presented in Table \ref{table: bs} and are later used in Section \ref{Section: para-full} to estimate the optimal division of computational resources when additional layers of parallelisation are implemented into the extrapolation method of the integrator.

\begin{center}
\begin{table}
\begin{tabular}{ |l|c|c|c|c| } 
 \hline
 $N_\mathrm{part}$ & $b_\mathrm{1}$ & $c_\mathrm{1}$ & $b_\mathrm{2}$ & $c_\mathrm{2}$\\ 
 \hline
 $264$  & $3.60$ & $5.10$  & $\times$ & $\times$\\ 
 $546$  & $3.24$ & $5.33$  & $3.16$   & $4.97$\\ 
 $1129$ & $5.03$ & $9.08$  & $4.02$   & $31.87$\\ 
 $2336$ & $6.50$ & $12.35$ & $6.09$   & $79.49$\\ 
 \hline
\end{tabular}
\caption{The force speed-up coefficients $b_\mathrm{i}$ and $c_\mathrm{i}$ obtained by fitting the data from Fig. \ref{fig: theorspeedup-2} using the function Eq. \eqref{eq: errorfun}. For the smallest particle number $N_\mathrm{part}=264$ fits beyond a single term were not profitable.}
\label{table: bs}
\end{table}
\end{center}

\subsection{Substep division parallelisation}\label{Section: para-div}

Solving initial value problems numerically for individual ordinary differential equations was long considered to be an inherently sequential process. However, numerical techniques employing extrapolation methods are an important exception to this rule (e.g. \citealt{Rauber1997, Korch2011} and references therein). As the N-body problem is an initial value problem for a coupled system of ordinary differential equations, it is possible to introduce another layer of parallelisation besides the parallel force loop computation into N-body codes which use extrapolation methods. To our best knowledge the only work studying orbital dynamics with a code including a parallelised extrapolation method is the study by \cite{Ito1997}. Unfortunately, this pioneering study has not received wide attention in the literature.

For the \mstar{} code implementation we use the Neville-Aitken algorithm (e.g. \citealt{Press2007}) for polynomial extrapolation. Error control is implemented as in Eq. \eqref{eq: gbs-criterion} by studying the relative difference of consecutive extrapolation results. We chose polynomials over rational functions for extrapolation as we have observed that using rational functions may occasionally lead to spurious extrapolation results even though the convergence criteria of Eq. \eqref{eq: gbs-criterion}  are fulfilled. We do not use the sophisticated recipes intended to optimise the amount of computational work per unit step in serial extrapolation methods, present in some GBS implementations \citep{Press2007,Hairer2008}.

\begin{figure}
\includegraphics[width=\linewidth]{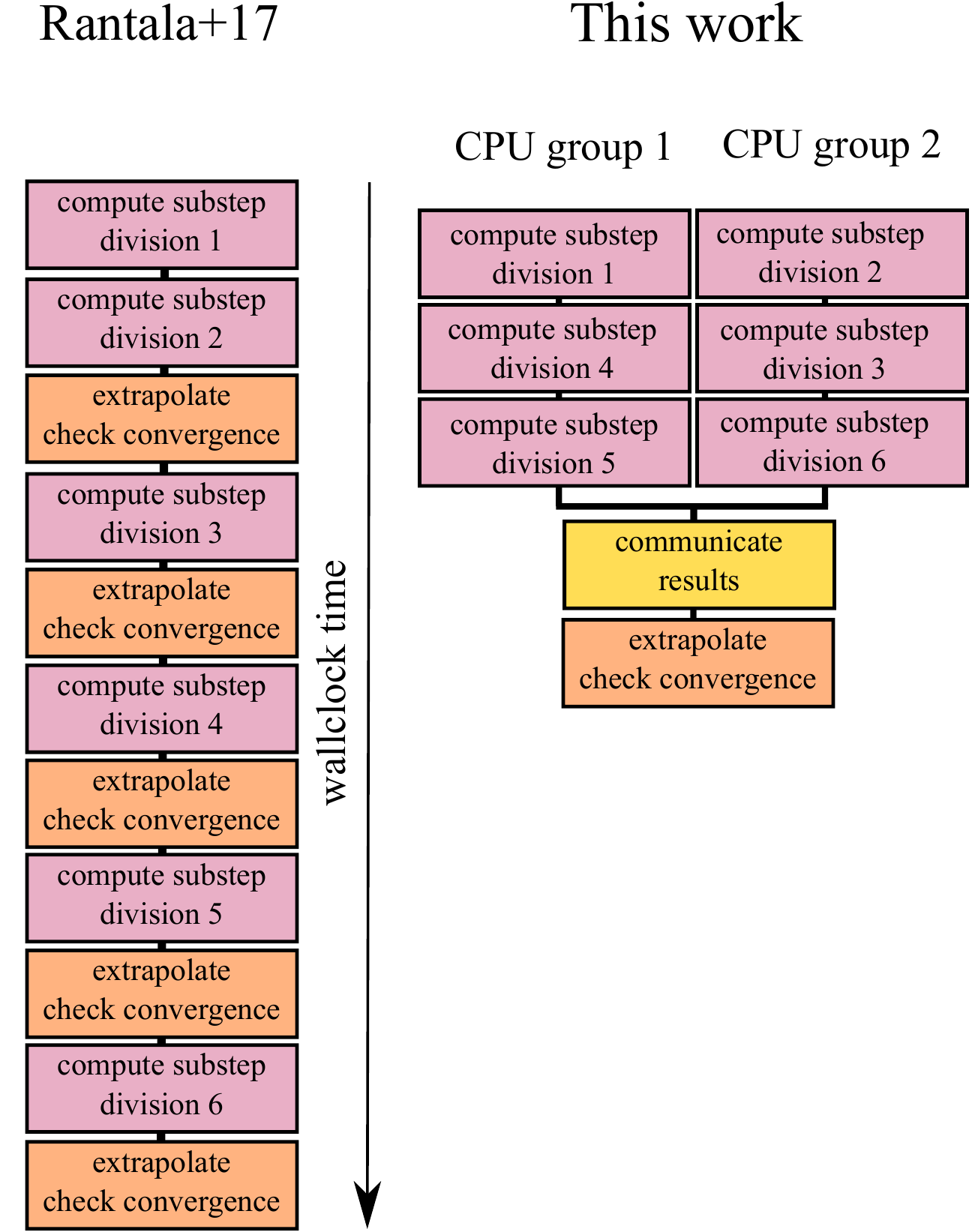}
\caption{The parallelisation strategy of the GBS extrapolation method without substep parallelisation as in our previous \archain{} implementation (left) and including it in our new \mstar{} code (right). In this example the extrapolation method converges after six substep divisions. The previous extrapolation method computes the different substep divisions, tries extrapolation and check convergence in a sequential manner. The parallelised extrapolation method computes the different substep divisions in parallel using different CPU groups, communicates the results to each node and then performs the extrapolation procedure.}
\label{fig: gbs}
\end{figure}

Implementing another parallelisation layer into an algorithmically regularised integrator begins with the observation that the computation of a single substep division in the GBS method is independent of the other required substep divisions. Thus, the different substep divisions can be integrated using different CPU groups in parallel \citep{Rauber1997}, after which the results of subdivisions are communicated and the actual extrapolation is performed. The Neville-Aitken polynomial extrapolation for all the dynamical variables can be parallelised as well. In this work we focus on the case of a single N-body system, but the method described here can be extended to integrate multiple independent N-body systems simultaneously. A simple example with two CPU groups and six substep divisions is illustrated in Fig. \ref{fig: gbs}. We label the number of CPU groups $N_\mathrm{div}$. As we use a single MPI task per CPU, the total number of CPUs, $N_\mathrm{CPU}$, the number of CPU groups $N_\mathrm{div}$, and the number of CPUs in force computation $N_\mathrm{force}$, are connected by the simple relation
\begin{equation}
    N_\mathrm{CPU} = N_\mathrm{div} \times N_\mathrm{force}.
\end{equation}

As stated in Section \ref{Section: gbsold}, the standard GBS method computes the substep divisions in sequential manner, calculating subsequent substep divisions until the results converge or the user-given maximum number of substep divisions $k_\mathrm{max}$ is reached. The parallelisation of the substep divisions requires that the number of substep divisions to be computed must be set by user in advance in our implementation. We call this fixed number of substep divisions $k_\mathrm{fix}$. We note that techniques to use a non-fixed $k_\mathrm{max}$ exist even with parallelisation \citep{Ito1997} but the simple approach with a fixed number of subdivisions has proven to be sufficient for our purposes. The optimal value for  $k_\mathrm{fix}$ depends on the GBS tolerance parameter $\etabs$ and the particular N-body system in question. Thus, numerical experimentation is needed for determining $k_\mathrm{fix}$. If $k_\mathrm{fix}$ is set to too low a value, the extrapolation method must shorten the time-step $H$ in order to reach convergence, increasing the running time of the code. On the other hand, if $k_\mathrm{fix}$ is too high, extra computational work is performed as the extrapolation would have converged with fewer substep divisions. However, this slow-down is somewhat compensated by the fact that longer time-steps $H$ can be taken with a higher $k_\mathrm{fix}$.

The number of CPU groups $N_\mathrm{div}$ can have values between $1\leq N_\mathrm{div} \leq k_\mathrm{fix}$ leaving $N_\mathrm{force} = N_\mathrm{CPU}/N_\mathrm{div}$ for parallelisation of the force computation. The individual substep divisions are divided into the $N_\mathrm{div}$ CPU groups with the following recipe. Initially, each CPU group has the computational load $C_\mathrm{i}=0$. Starting from the substep division with the highest number of substeps, i.e. $\max(n_\mathrm{k}) = 2k_\mathrm{fix}$, we assign the substep divisions one by one into the CPU group which has the lowest computational load $C_\mathrm{i}$ at that moment until no divisions remain. If there are several CPU groups with the same computational load we select the first one, i.e. the CPU group with the lowest CPU group index in the code.
\begin{figure}
\includegraphics[width=\linewidth]{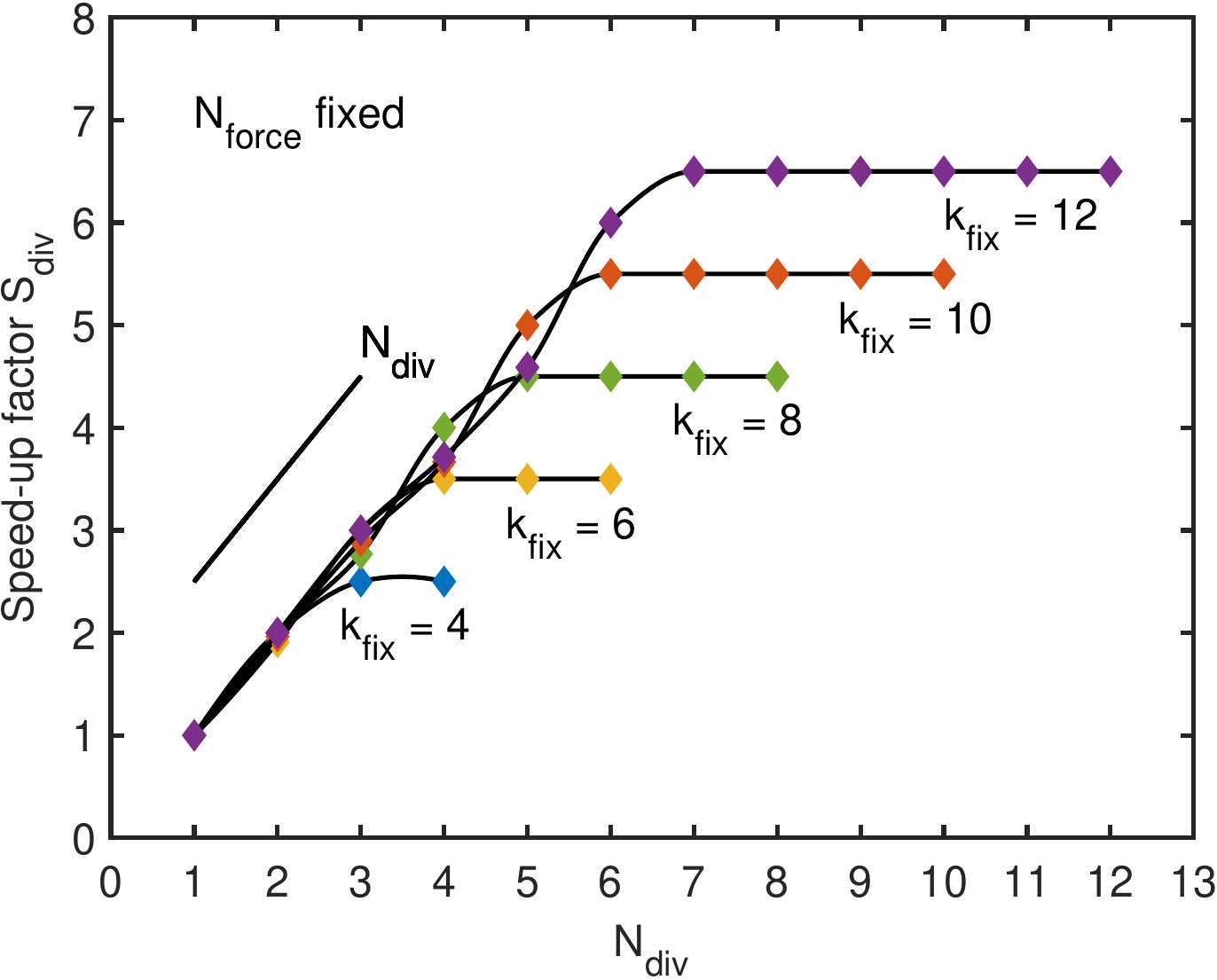}
\caption{The speed-up factor $S_\mathrm{div}$ for the parallelised substep divisions (symbols) and their interpolated continuous counterparts (solid lines) as a function of the number of substep divisions $k_\mathrm{fix}$. The interpolants are only shown to clarify the visualisation as there are no groups with non-integer $k_\mathrm{max}$. We see that $S_\mathrm{div} = N_\mathrm{div}$ until $N_\mathrm{div} = k_\mathrm{fix}/2$ after which the speed-up factor attains a constant value of $S_\mathrm{div}^\mathrm{max} = (k_\mathrm{fix}+1)/2$.}
\label{fig: theorspeedup-1}
\end{figure}
Keeping $N_\mathrm{force}$ fixed we define the parallel substep division speed-up $S_\mathrm{div}$ as
\begin{align}\label{eq: speedup-div}
\begin{split}
    &S_\mathrm{div}(N_\mathrm{div}) = \frac{T_\mathrm{serial}}{T_\mathrm{parallel}(N_\mathrm{div})} = 
    \frac{ \sum_{k=1}^{k_\mathrm{fix}} n_k }{ \max_i C_i }\\ & =\frac{\sum_{k=1}^{k_\mathrm{fix}} n_k }{ \max_i \left( \left[ \sum_j^{N_\mathrm{i}} n_{k_j} \right]_i  \right) }
\end{split}
\end{align}
If $N_\mathrm{div} = 1$ there is no speed-up in the force computation an the running time is the same as in Eq. \eqref{eq: force-wallclock}. When $N_\mathrm{div} > 1$ there is a wall-clock time speed-up as the force computation is divided into multiple CPU groups. With $N_\mathrm{div} = k_\mathrm{fix}$ we have $S_\mathrm{div} = (k_\mathrm{fix}+1)/2$ assuming the Deuflhard sequence from Eq. \eqref{eq: seq-deuflhard}.

Now we can compute the speed-up factor $S_\mathrm{div}$ once $k_\mathrm{fix}$ and $N_\mathrm{div}$ are set. The computed results are presented in Fig. \ref{fig: theorspeedup-1}. The speed-up factor follows the line $S_\mathrm{div} = N_\mathrm{div}$ until the point $N_\mathrm{div} = k_\mathrm{fix}/2$ is reached, after which the $S_\mathrm{div}(N_\mathrm{div})$ rapidly flattens into the constant value of $S_\mathrm{div} = (k_\mathrm{fix}+1)/2$. Thus, the maximum reasonable number of CPU groups is $N_\mathrm{div}^\mathrm{max} = \ceil{k_\mathrm{fix}/2}$ in which we use the ceiling function $\ceil{\cdot}$.

\subsection{Speed-up with full parallelisation}\label{Section: para-full}

\begin{figure*}
\includegraphics[width=\textwidth]{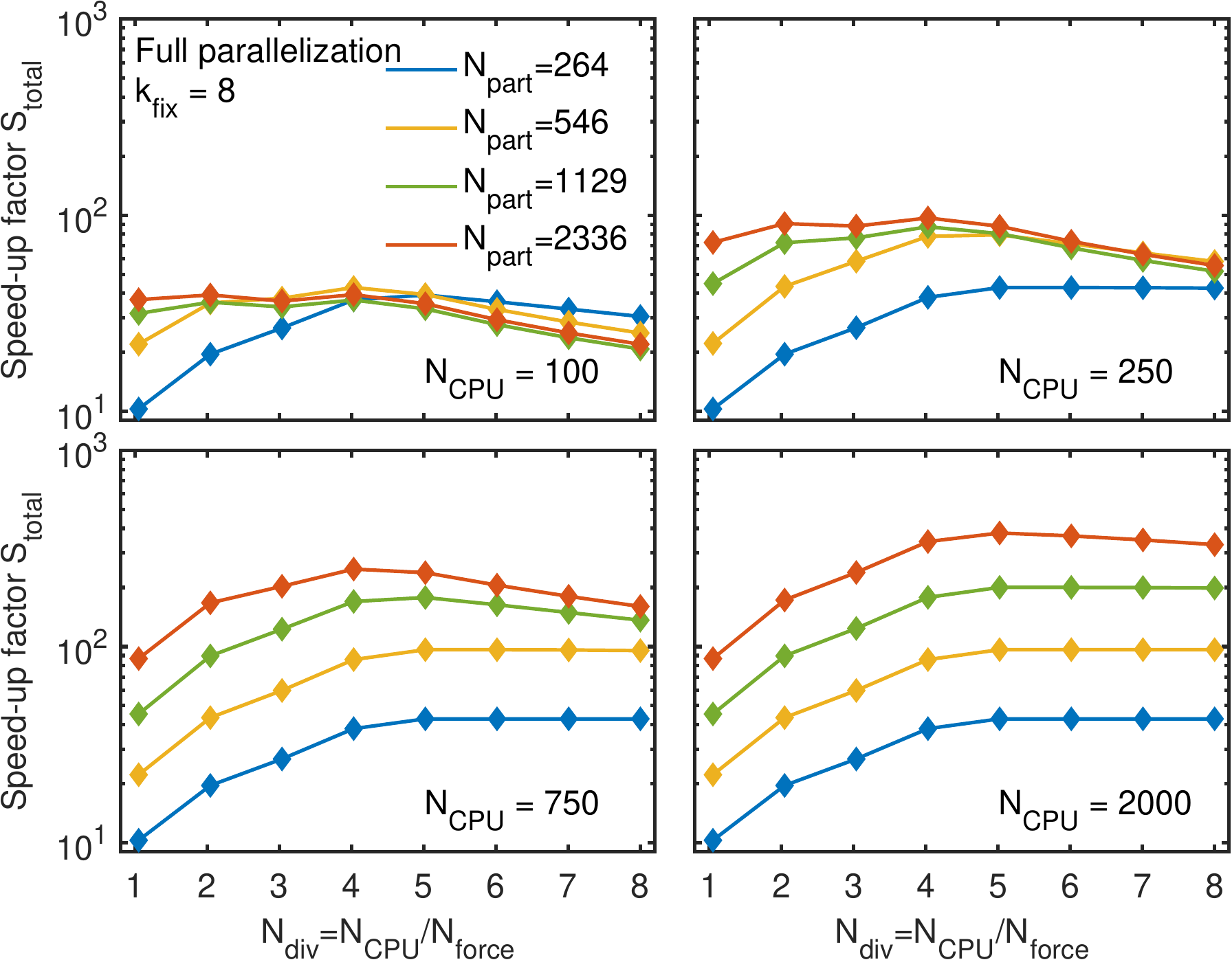}
\caption{An example of finding the optimal division of computational resources between the force and substep parallelisation. The substep part of the speed-up factor can be computed analytically while the force parallelisation part is estimated by using simple numerical tests as explained in the text. Starting from the top-left corner, the four panels have increasing CPU numbers of $N_\mathrm{CPU}=100$, $N_\mathrm{CPU}=250$, $N_\mathrm{CPU}=750$ and $N_\mathrm{CPU}=1000$. Each panel shows the speed-up factor $S_\mathrm{total}$ as a function of the number of CPU groups $N_\mathrm{div}$ for four different particle numbers of $N_\mathrm{part}=264$ (blue line), $N_\mathrm{part}=546$ (yellow), $N_\mathrm{part}=1129$ (green line) and $N_\mathrm{part}=2336$ (red line). The maximum speed-up factor $S_\mathrm{total}^\mathrm{max}$ is typically found near $N_\mathrm{div} = k_\mathrm{fix}/2$. The corresponding number of CPUs for the force computation is obtained by using the relation $N_\mathrm{force} = N_\mathrm{CPU}/N_\mathrm{div}$.}
\label{fig: theorspeedup-3}
\end{figure*}

Now we are ready to combine the force loop and the substep division layers of parallelisation. The primary advantage of using two layers of parallelisation compared to the simple force loop computation parallelisation is that we can efficiently use more MPI tasks to speed up the \mstar{} integrator. Without the subdivision parallelisation it is not reasonable to use more than $N_\mathrm{force} \approx 0.1\times N_\mathrm{part}$ MPI tasks as shown in Section \ref{Section: para-force}. With the subdivision parallelisation included, the maximum reasonable CPU (or task) number becomes $N_\mathrm{CPU} \approx 0.05\times k_\mathrm{fix} N_\mathrm{part} = 0.4 \times N_\mathrm{part}$ with the typical value of $k_\mathrm{fix} = 8$. This is the value of $k_\mathrm{fix}$ we use in the simulations of this study.

Next we estimate how the computational resources should be divided to ensure the maximum total speed-up factor $S_\mathrm{total}$ if the number of CPUs, $N_\mathrm{CPU}$, and thus MPI tasks, is fixed. The values of $N_\mathrm{part}$ and $k_\mathrm{fix}$ are assumed to be fixed as well. The optimal division of computational resources corresponds to finding the maximum of the function
\begin{align}
\begin{split}
&S_\mathrm{total}(N_\mathrm{force},N_\mathrm{div}) = S_\mathrm{force}(N_\mathrm{force}) \times S_\mathrm{div}(N_\mathrm{div})\\
& = S_\mathrm{force}(N_\mathrm{CPU}/N_\mathrm{div}) \times S_\mathrm{div}(N_\mathrm{div})
\end{split}
\end{align}
with the constraints $N_\mathrm{force}, N_\mathrm{div} \in \mathbb{N}$ and $N_\mathrm{CPU}=N_\mathrm{force} \times N_\mathrm{div}$. For arbitrary $N_\mathrm{CPU}$ there are typically only a few solutions. For the force computation speed-up $S_\mathrm{force}$ we need to use the approximate methods i.e. the fitting function Eq. \eqref{eq: errorfun} and its coefficients $b_\mathrm{i}$ and $c_\mathrm{i}$ from Table \ref{table: bs}. The substep division speed-up factor $S_\mathrm{div}$ can be exactly estimated by using Eq. \eqref{eq: speedup-div}.

In Fig. \ref{fig: theorspeedup-3} we present the speed-up factor $S_\mathrm{total}$ for four different particle numbers and four different values for $N_\mathrm{CPU}$. We set $k_\mathrm{max} = 8$ for each of the $16$ combinations of the particle number and the number of CPUs. For a fixed particle number the total speed-up factor $S_\mathrm{total}$ increases until $N_\mathrm{CPU} \sim 0.4 \times N_\mathrm{part}$. We find that the maximum of $S_\mathrm{total}$ is typically located near $\ceil{N_\mathrm{div}/2}$, which corresponds to finding the optimal $N_\mathrm{force}$ around $N_\mathrm{CPU}/(2N_\mathrm{div})$.

However, we find that the best strategy for finding the optimal pair $(N_\mathrm{force},N_\mathrm{div})$ is to relax the requirement of having a pre-set value for $N_\mathrm{CPU}$. One computes the values for $S_\mathrm{total}$ for all the integer pairs $(N_\mathrm{force},N_\mathrm{div})$ satisfying $1 \leq N_\mathrm{force} \leq \ceil{0.1 \times N_\mathrm{part}}$ and $1 \leq N_\mathrm{div} \leq \ceil{k_\mathrm{fix}/2}$. The location of the maximum value of $S_\mathrm{total}$ determines which values of $N_\mathrm{force}$ and $N_\mathrm{div}$, and thus also $N_\mathrm{CPU}$, should be used. Additional constraints such as the number of CPUs per supercomputer node should also be taken into account i.e. $N_\mathrm{CPU}$ should be a multiple of this number.

\begin{figure}
\includegraphics[width=\linewidth]{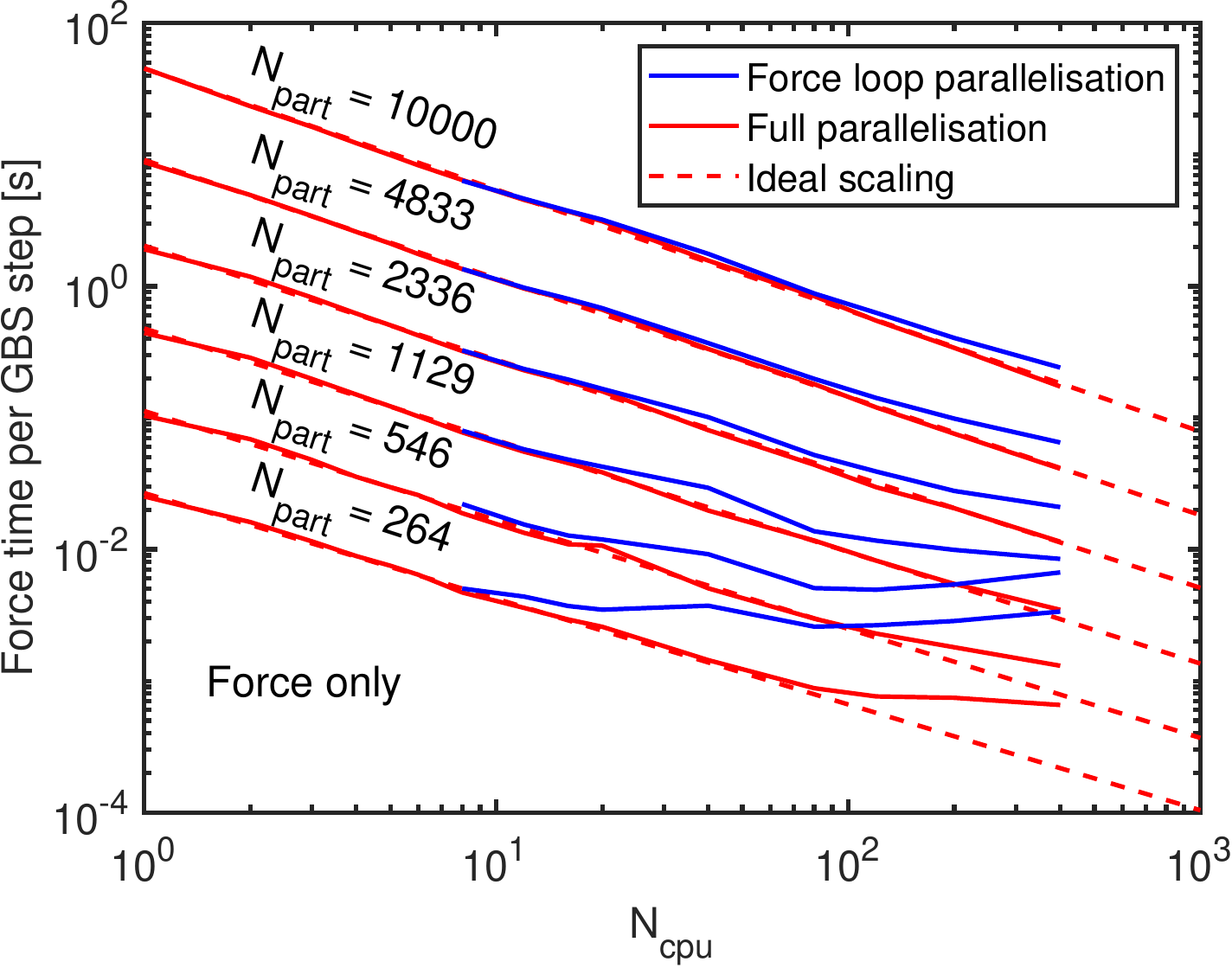}
\caption{The strong scaling test of the force loop parallelised (solid blue line) and the fully parallelised (solid red line) force computation algorithms. The serial running time per GBS step corresponds to Eq. \eqref{eq: force-wallclock}. The dashed red line shows the ideal scaling behaviour of the codes. The fully parallelised algorithm follows the ideal scaling behaviour up to $N_\mathrm{CPU} \sim 0.5 \times N_\mathrm{part}$ while the loop parallelised algorithm begins to deviate from the ideal scaling law already with roughly ten times smaller CPU numbers.}
\label{fig: forcescaling}
\end{figure}

Finally we present the results of a strong scaling test of our force calculation algorithms in Fig. \ref{fig: forcescaling}. In a strong scaling test the problem size remains fixed while the number of CPUs is increased. We examine both the force loop parallelised version and the code with full parallelisation. We see that the force calculation algorithm with full parallelisation follows the ideal scaling behaviour to higher CPU numbers than the force loop parallelised version. With the fully parallelised force computation one can use CPU numbers approximately up to $N_\mathrm{CPU} \sim 0.5\times N_\mathrm{part}$ before the scaling begins to deviate from the ideal case. Including only the force loop parallelisation the scaling behaviour becomes non-ideal with roughly ten times smaller $N_\mathrm{CPU}$. We tested the force algorithms up to $N_\mathrm{CPU}=400$ and all fully parallelised tests with particle numbers $N_\mathrm{part} \gtrsim 10^3$ followed ideal scaling. The total speed-up factors $S_\mathrm{total}$ are consistent with our estimations in this Section.

\section{Code accuracy: few-body tests} \label{Section: code accuracy}

\subsection{Eccentric Keplerian binary}

\begin{figure*}
\includegraphics[width=\textwidth]{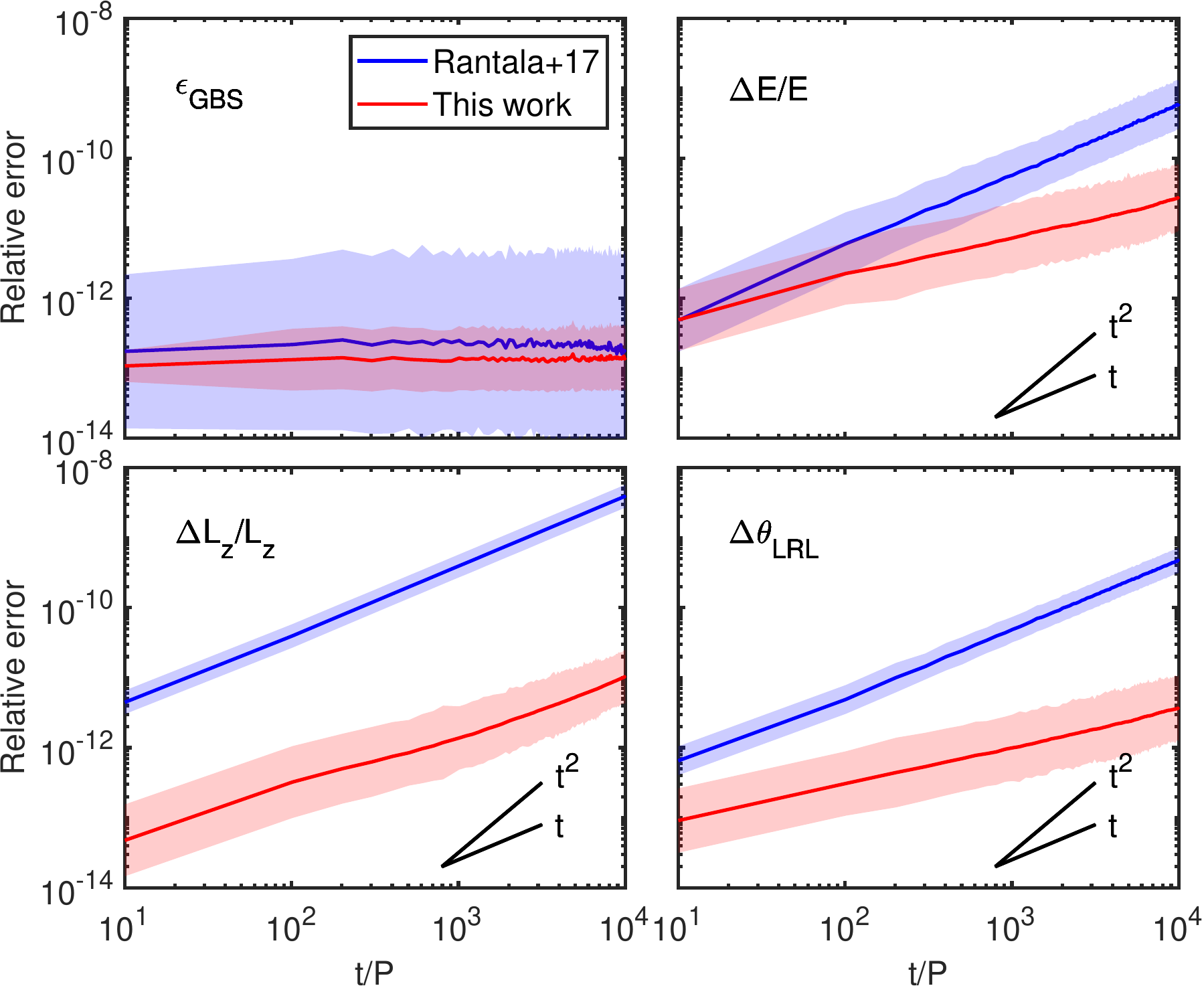}
\caption{Results of the SMBH binary simulations with our \archain{} implementation (blue) and \mstar{} (red). Starting from the top left panel, the four panels show the maximum extrapolation error $\epsilon_\mathrm{GBS}$ after an accepted step, the relative error in energy $E$ and angular momentum $L_\mathrm{z}$ and the rotation angle of the Laplace--Runge--Lenz vector $\theta_\mathrm{LRL}$. The error regions depict a single standard deviation. The lower scatter in the maximum extrapolation error in \mstar{} indicates that the code does not exceedingly increase the step-size after a successful step which would lead to divergence and step split during the next step. Our the new code clearly performs better than our \archain{} implementation with the numerical errors being smaller throughout.}
\label{fig: binary1}
\end{figure*}

Next we demonstrate the numerical accuracy of our \mstar{} integrator by studying the standard Keplerian two-body problem by comparing the results to the analytical solution and to our \archain{} code \citep{Rantala2017}. In the following Sections we also runs tests with two additional three-body setups.

All the simulation runs of this study are run on the FREYA\footnote{www.mpcdf.mpg.de/services/computing/linux/Astrophysics} cluster of the Max Planck Computing and Data Facility (MPCDF). Each computation node of FREYA contains two Intel Xeon Gold 6138 CPUs totalling $40$ cores per computation node. However, in the context of this article we refer to these core units as CPUs.

The Keplerian two-body problem is completely described by its six integrals of motion. As the final integral, the periapsis time, can be arbitrarily chosen we only need five integrals of motion to describe the orbit. 
The first conserved quantity is the energy $E$ of the two-body system defined as
\begin{equation}
    E = \frac{1}{2} \mu  \norm{\vect{v}}^2 - \frac{G \mu M}{\norm{\vect{r}}}
\end{equation}
in which $M=m_\mathrm{1}+m_\mathrm{2}$, $\mu = m_\mathrm{1}m_\mathrm{2}/M$ and $\vect{r}$ and $\vect{v}$ are the relative position and velocity vectors, respectively. The energy of the two-body system uniquely defines its semi-major axis $a$ as
\begin{equation}
    a = -\frac{G \mu M}{2 E}
\end{equation}
Next, the conserved angular momentum vector $\vect{L}$ defined as
\begin{equation}
    \vect{L} = \mu \vect{r} \times \vect{v}.
\end{equation}
Together $E$ and $\vect{L}$ determine the orbital eccentricity $e$ of the two-body system as
\begin{equation}
e = \left( 1 + \frac{2 E L^2}{G \mu^3 M^2} \right)^{1/2}.
\end{equation}
Finally, we have the constant Laplace--Runge--Lenz vector
\begin{equation}
    \vect{A} = \mu \vect{v} \times \vect{L} - G M\vect{\hat{r}}
\end{equation}
in which $\hat{r}$ = $\vect{r}/\norm{\vect{r}}$. The Laplace--Runge--Lenz vector lies in the orbital plane of the two-body system pointing towards the periapsis. As we have now in total seven conserved quantities and only five integrals are required the conserved quantities cannot be independent. The first relation is simply $\vect{A} \cdot \vect{L} = 0$ while the non-trivial second relation reads $e = \norm{\vect{A}}/(GM)$, connecting both the energy $E$ and the norm of the angular momentum vector $L$ to the norm of $\vect{A}$.

It is convenient to study the accuracy of a numerical integrator by observing the behaviour of $E$, $\vect{L}$ and $\vect{A}$ during a simulation run. Symplectic integrators such as our chained leapfrog typically conserve quadratic invariants such as the angular momentum exactly and the energy approximately but with no long-term secular error growth \citep{Hairer2006}. However, the Laplace--Runge--Lenz vector is a third order invariant, and its conservation is not guaranteed. Thus the orbit can precess in its orbital plane (e.g. \citealt{Springel2005}). This makes the rotation angle of the Laplace--Runge--Lenz vector 
\begin{equation}
    \theta_\mathrm{LRL} = \arctan(A_\mathrm{y}/A_\mathrm{x})
\end{equation}
a very suitable probe for testing the accuracy of an integrator.

We perform a series of two-body simulations both with our \mstar{} integrator and our \archain{} implementation. For the tests in this Section, serial code implementations are used. We initialise $360$ equal-mass SMBH binaries with $M = 2\times10^9 M_\odot$, $a=2$ pc and $e=0.9$. We orient the initial binaries in a random orientation in space in order to have a sample of binaries with initially the same integrals of motion but with differing numerical errors during the simulation. We run the binary simulations for $T = 10^4\times P$ in which $P$ is the Keplerian orbital period of the binary. The GBS tolerance is set to $\etabs = 10^{-12}$. We always use $k_\mathrm{fix}=8$ substep divisions in the serial GBS procedure.

\begin{figure}
\includegraphics[width=\linewidth]{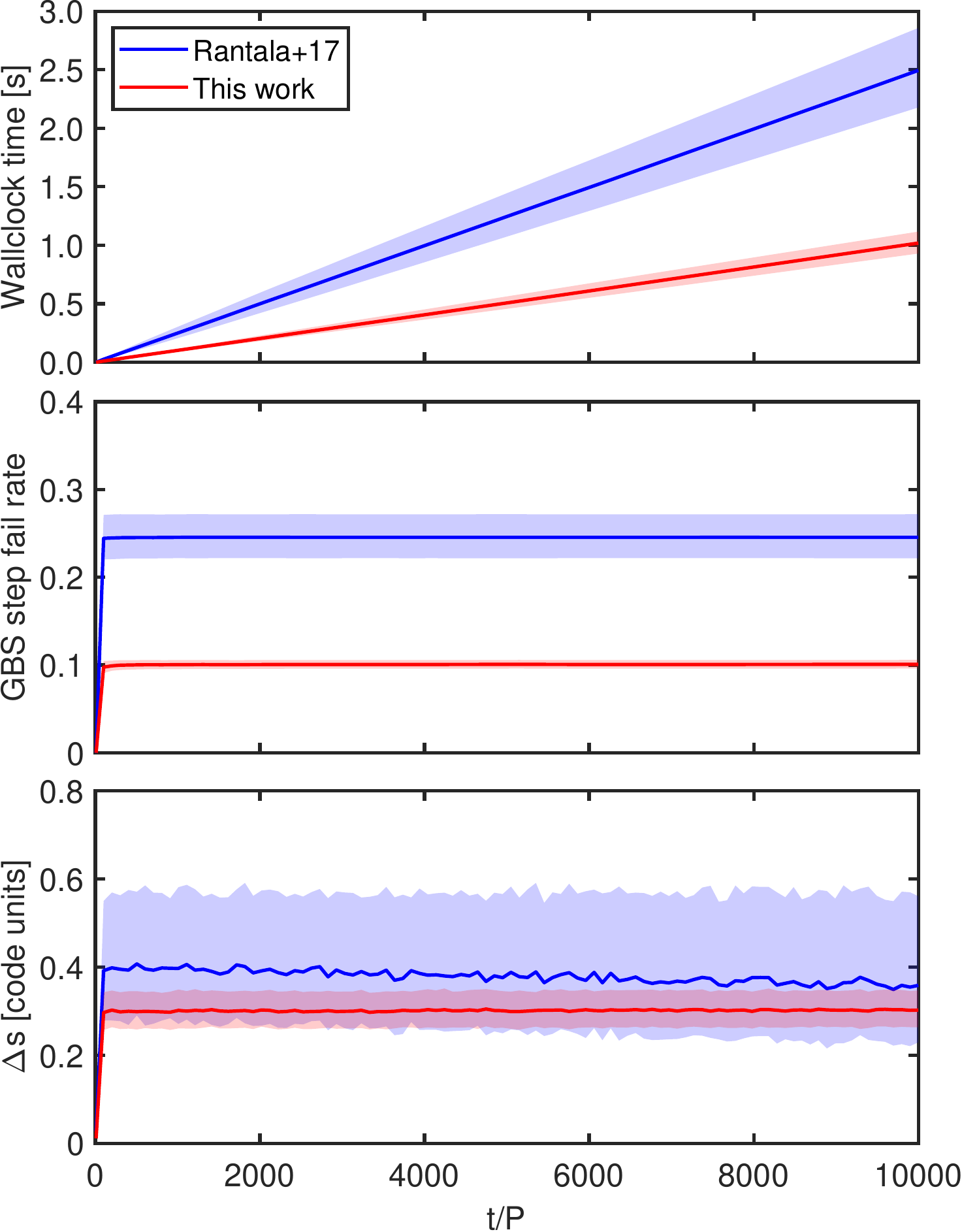}
\caption{Additional results of the binary simulations with our previous \archain{} implementation (blue) and the new code (red). The top panel presents the wall-clock times spent by the integrators with the new implementation being $2$--$3$ times faster than the old one. The middle and the bottom panels show the GBS step fail rate and the GBS step size $H$ in fictitious time $\Delta s$. These panels confirm that the new \archain{} implementation is faster in two-body tests due to its factor of $\sim 2.5$ smaller GBS step fail rate.}
\label{fig: binary2}
\end{figure}

The results of the binary simulations are presented in Fig. \ref{fig: binary1} and Fig. \ref{fig: binary2}. The panels of Fig. \ref{fig: binary1} illustrate the relative error of the energy and the norm of the angular momentum vector as well as the absolute rotation angle of the Laplace--Runge--Lenz vector. In addition we show the maximum GBS error $\epsilon_\mathrm{GBS}$ after convergence in each step. Fig. \ref{fig: binary2} in turn presents the elapsed wall-clock time, the GBS step fail rate and the length of the fictitious time-step during the simulations.

The results of the binary simulations systematically show that the new \mstar{} implementation conserves the orbital energy $E$, angular momentum $\vect{L}$ and the Laplace--Runge--Lenz vector $\vect{A}$ $\sim 1 \text{--} 2$ orders of magnitude better than our old \archain{} implementation. In addition, the new code is faster than the old code by a factor of few. The difference in the code speed originates from the fact that the GBS recipe of \cite{Hairer2008} that we are using in our \archain{} implementation optimises the computational work per unit step and also aggressively attempts longer steps $H$ after convergence. This leads to a large number of failed GBS steps, slowing down the code in the test. However, the implementation is not very transparent and thus it is somewhat difficult to point to exactly where the speed and accuracy differences originate compared to our own implementation of the extrapolation algorithm in \mstar.

\subsection{Pythagorean three-body problem}

The Pythagorean three-body problem \citep{Burrau1913, Szebehely1967} is a famous zero angular momentum setup to test integrators and to study chaotic and orderly motion in a relatively simple gravitating system \citep{Aarseth1994, Valtonen2006}. Three SMBHs with masses of $M_\mathrm{1} = 3\times10^8 M_\odot$, $M_\mathrm{2} = 4\times10^8 M_\odot$ and $M_\mathrm{3} = 5\times10^8 M_\odot$ are placed at the corners of a right-angled triangle with side lengths of $r_\mathrm{13} = 30$ pc, $r_\mathrm{12} = 40$ pc and $r_\mathrm{23} = 50$ pc i.e. in such a manner that the least massive SMBH is opposite to the shortest side and so on. Initially all velocities are set to zero. Once the simulation run is started, the three bodies experience a series of complicated interactions finally resulting in the ejection of the least massive $M_\mathrm{1}$ body while the remaining two SMBHs form a bound binary recoiling to the opposite direction.

\begin{figure}
\includegraphics[width=\linewidth]{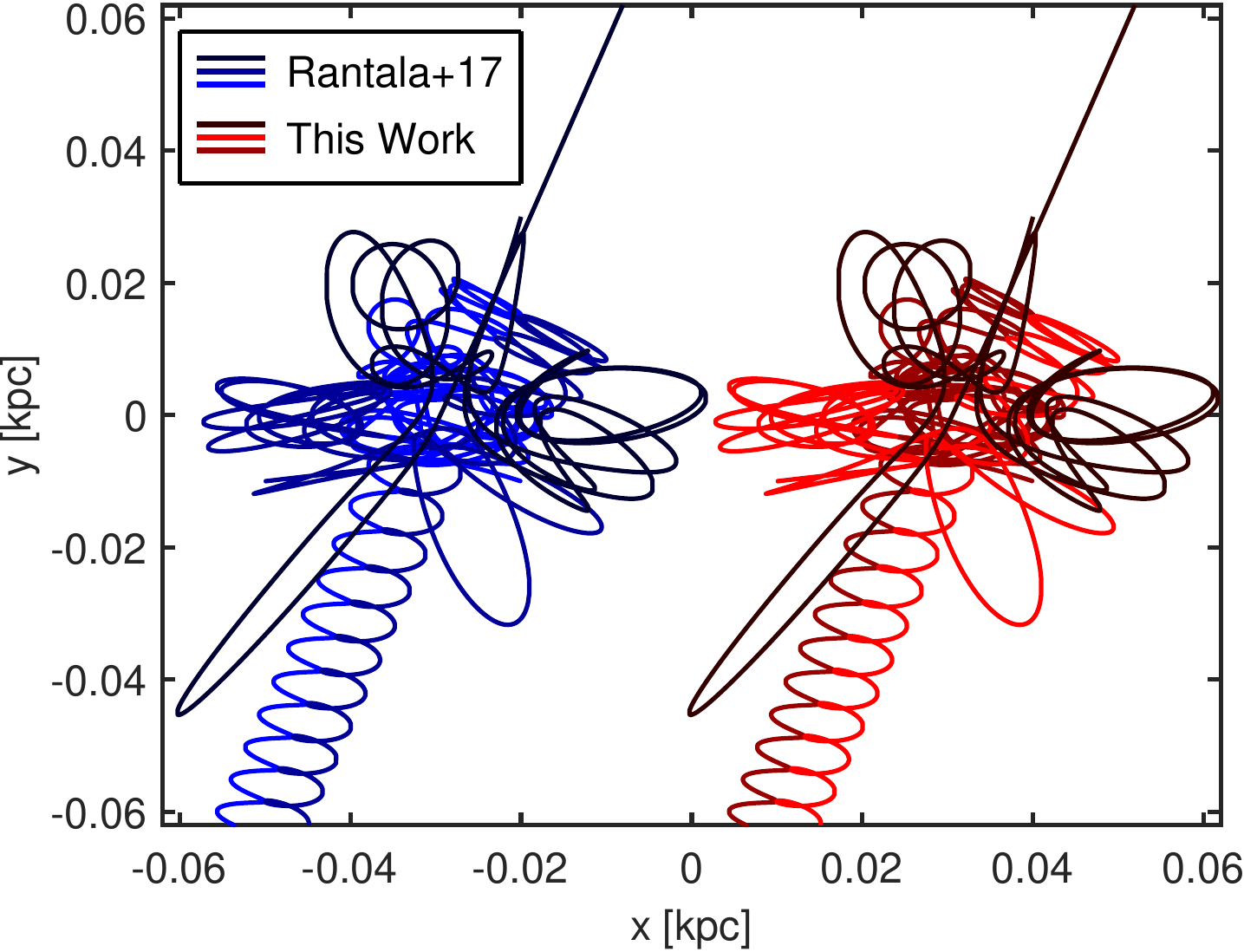}
\caption{The general overview of the orbits of the three bodies in the Pythagorean three-body problem. Initially the three bodies are gravitationally bound, but after a series of complicated interactions the least massive body (black line) is ejected while the other two bodies (red and blue) form a bound binary recoiling at the opposite direction. By eye, there are no noticeable differences in the results with the two integrators.}
\label{fig: pyt-1}
\end{figure}

The final outcome of the system can be parametrised by the orbital elements ($a_\mathrm{2,3}$,$e_\mathrm{2,3}$) of the formed binary and the escape direction $\beta_\mathrm{1}$. If the initial triangle setup is oriented as in Fig. 1 of \cite{Aarseth1994}, the escape angle becomes

\begin{equation}
    \beta_\mathrm{1} = \arctan(y_\mathrm{1}/x_\mathrm{1})
\end{equation}
in which the subscript refers to the least massive SMBH. The system is extremely sensitive to the initial conditions and the numerical accuracy of the used integrator, thus providing an ideal test setup for demonstrating that our old \archain{} and the new \mstar{} implementations yield the same final results.

We perform the integration with the same code parameters as the two-body tests. We show the orbits of the three SMBHs in the Pythagorean three-body problem in Fig. \ref{fig: pyt-1} both with the old \archain{} and the new \mstar{} integrator. The overall behaviour of the system is as expected from the literature results: the least massive body becomes unbound and the binary of the two remaining bodies recoils in the opposite direction. At this level of scrutiny there are no noticeable differences between the two integrator implementations.

\begin{figure}
\includegraphics[width=\linewidth]{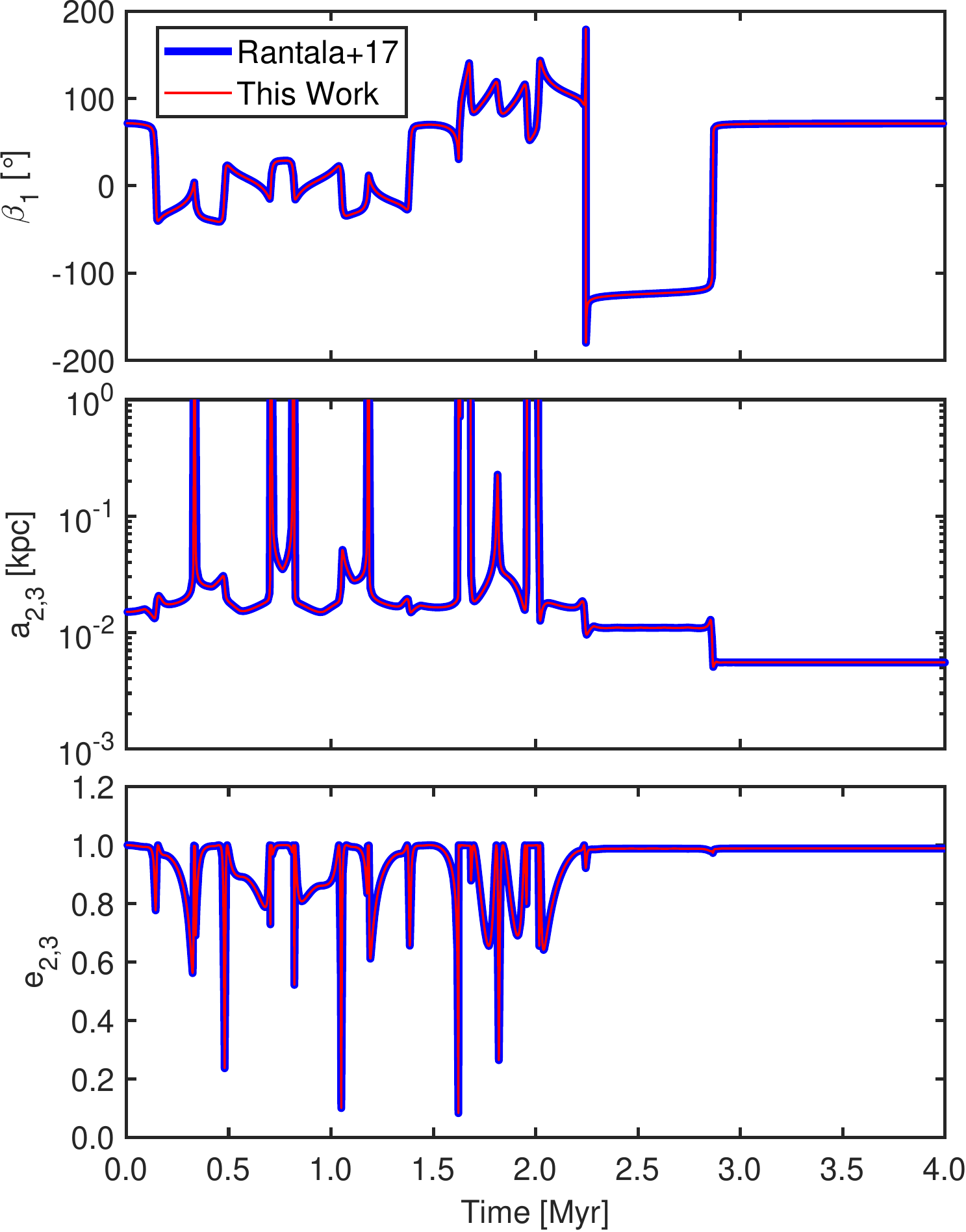}
\caption{The three variables $\beta_\mathrm{1}$,$a_\mathrm{2,3}$ and $e_\mathrm{2,3}$ parameterising the outcome of the Pythagorean three-body problem as described in the main text. The results of the \mstar{} and the \archain{} integrator always agree within a relative factor of $10^{-4}$.}
\label{fig: pyt-2}
\end{figure}

The escape angle $\beta_\mathrm{1}$ as well as the orbital elements of the pair of two most massive bodies are presented in Fig. \ref{fig: pyt-2}. After a period of complicated gravitational dynamics the values of $\beta_\mathrm{1}$, $a_\mathrm{2,3}$ and $e_\mathrm{2,3}$ settle to their final values as the motion of the system becomes ordered after the escape of the least massive SMBH. Both the \archain{} and the \mstar{} integrator implementations provide the results $\beta_\mathrm{1} \approx 71.4^\circ$, $a_\mathrm{2,3} \approx 5.5$ pc and $e_\mathrm{2,3}\approx 0.99$ with a relative difference of only $\sim 10^{-4}$ in each value. Due to the extreme sensitivity of the Pythagorean three-body problem to numerical errors during integration we conclude that the two integrators produce the same results within an accuracy sufficient for our purposes. These final results agree also very well with the literature values \citep{Szebehely1967, Aarseth1994}.

\subsection{Lidov--Kozai oscillations}

The Lidov--Kozai mechanism \citep{Lidov1962, Kozai1962} is a widely studied dynamical phenomenon present in a family of hierarchical three-body systems. The mechanism has a large number of applications in dynamical astronomy reaching from dynamics of artificial satellites to systems of supermassive black holes \citep{Naoz2016}. An inner binary consisting of a primary and a secondary body is perturbed by a distant orbiting third body. The inner binary and the perturber form the outer binary. The time-varying perturbation causes the argument of pericenter of the secondary body to oscillate around a constant value. Consequently, the eccentricity and the inclination of the inner binary with respect to the orbital plane of the outer binary oscillate as well. The time-scale of the oscillations exceeds by far the orbital periods of the inner and outer binaries.

In the limit of the secondary body being a test particle the quantity
\begin{equation}
l_\mathrm{z} = \sqrt{1-e_\mathrm{2}^2}\cos{i_\mathrm{2}}
\end{equation}
is conserved. Here the subscripts of the orbital elements refer to the secondary body with respect to the primary body. The Lidov--Kozai oscillations are present in the three-body system if the inclination $i_\mathrm{0}$ of the secondary exceeds the critical value $i_\mathrm{crit}$ defined as
\begin{equation}
    i_\mathrm{crit} = \arccos{\left(\sqrt{\frac{3}{5}}\right)}
\end{equation}
which is approximately $i_\mathrm{crit} \approx 39.2^\circ$. The maximum achievable eccentricity $e_\mathrm{max}$ depends only on the initial inclination $i_\mathrm{0}$ as
\begin{equation}
    e_\mathrm{max} = \sqrt{ 1-\frac{5}{3} \cos{}^2 i_\mathrm{0}}.
\end{equation}

We set up a hierarchical three-body system with masses of $M_\mathrm{1} = M_\mathrm{3} = 10^9 M_\odot$ and $M_\mathrm{2} = 10^3 M_\odot$ using the following orbital parameters. The outer binary is circular ($e_\mathrm{outer}=0$) with a semi-major axis of $a_\mathrm{outer} = 20$ pc. The inner binary is initially almost circular ($e_\mathrm{inner}=10^{-3}$) and has a semi-major axis of $a_\mathrm{inner} = 2$ pc. The orbital plane of the secondary is initially inclined $i_\mathrm{0} = 80^\circ$ with respect to the orbital plane of the outer binary, exceeding the critical inclination $i_\mathrm{crit}$ so the system exhibits Lidov--Kozai oscillations. The test particle approximation predicts the maximum eccentricity of $e_\mathrm{max} \approx 0.975$ for the system.

\begin{figure}
\includegraphics[width=\linewidth]{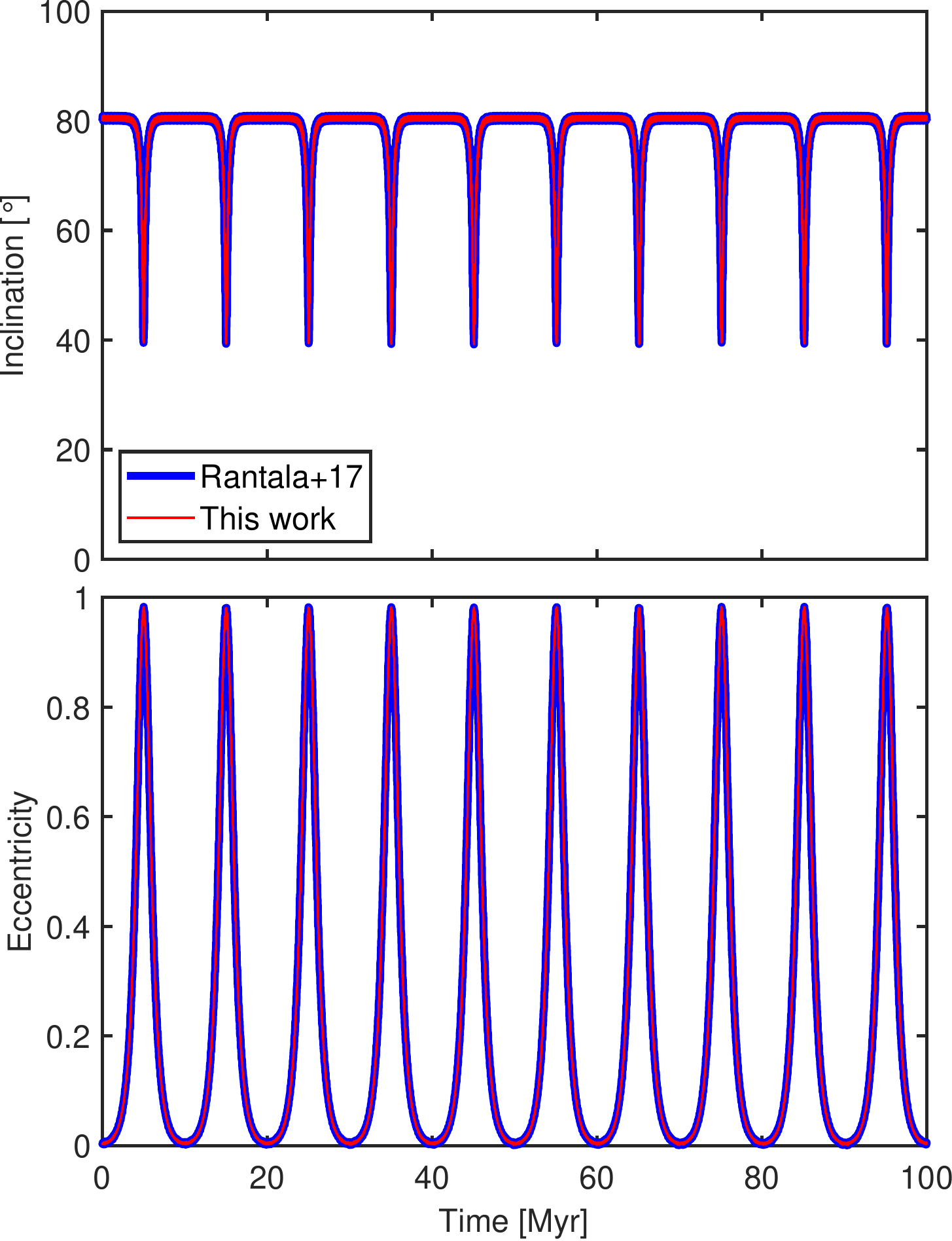}
\caption{The Lidov--Kozai mechanism test with the \archain{} (blue line) and the \mstar{} integrator (red line). The hierarchical three-body system shows strong oscillations both in the inclination (top panel) and the eccentricity (bottom panel) of the secondary body. The results of the two integrators agree very well with each other and analytical estimates as described in the text.}
\label{fig: kozai}
\end{figure}

We simulate the evolution of the three-body system for $100$ Myr using both the \archain{} and \mstar{} integrators. The integrator accuracy parameters are identical to the ones in the previous Section. The oscillations of eccentricity and inclination of the secondary during the simulation are presented in Fig. \ref{fig: kozai}. The system experiences roughly ten strong oscillations in $100$ Myr reaching a maximum eccentricity of $e_\mathrm{max} = 0.983$. The minimum inclination during the oscillations is very close to the critical value of $i_\mathrm{crit} \approx 39.2^\circ$. The system evolves practically identically when run with the old \archain{} and the new \mstar{} integrator. The relative difference of the value of $e_\mathrm{max}$ with the two integrators is only of the order of $10^{-8}$.

\section{Code scaling and timing: N-body tests} \label{Section: code scaling}

\subsection{N-body initial conditions}\label{Section: ic}

We construct gravitationally bound clusters of equal-mass point particles in order to perform code timing tests. We use $20$ different particle numbers $N_\mathrm{part}$, selected logarithmically between $N_\mathrm{part}=10^1$ and $N_\mathrm{part}=10^4$ particles with three different random seeds for each run, totalling $60$ cluster initial conditions. The particle positions are drawn from the spherically symmetric Hernquist sphere \citep{Hernquist1990} with a density profile of
\begin{equation}
\rho(r) = \frac{M}{2 \pi} \frac{a_\mathrm{H}}{r(r+a_\mathrm{H})^3},
\end{equation}
where $M$ is the total mass of the system and $a_\mathrm{H}$ its scale radius. We set $M = 10^7 M_\odot$ and $a_\mathrm{H}$ in such a manner that the half-mass radius of the system equals $r_\mathrm{1/2} = 10$ pc. The particle velocities are sampled from the Hernquist density-potential pair using the Eddington's formula technique (e.g. \citealt{Binney2008}).

Even though we do not intentionally include primordial binaries in our cluster construction setup, binaries may still form when sampling particle positions and velocities. A binary is considered hard if its binding energy exceeds the average kinetic energy of a particle in the cluster, i.e.
\begin{equation}
\frac{G \mu M}{2 a} \gtrsim \frac{1}{2} m \sigma^2,
\end{equation}
where $m$ is the mass of a single particle and $\sigma$ is the velocity dispersion of the cluster. While our \mstar{} integrator can easily compute the dynamics of hard binaries, the possible existence of such binaries is problematic for the N-body timing tests. This is because an integrator using the logarithmic Hamiltonian (or equivalent) time transformation can propagate physical time only for an amount $\Delta t$ per one step, where $\Delta t$ is of the order of the orbital period $P$ of the hardest binary in the cluster, defined as
\begin{equation}
P = 2 \pi\left( \frac{a^3}{GM} \right)^{1/2}
\end{equation}
by the Kepler's third law. Consequently, the total running time of the simulation will scale as $\Delta t^{-1} \propto a^{-3/2}$. This is very inconvenient as the clusters with the same $N_\mathrm{part}$ but a different binary content may have a very large scatter in their simulation times up to several orders of magnitude. Thus, we exclude all clusters which contain even a single hard binary and generate a new cluster until we have $60$ clusters in total. For the same reason we do not include a single heavy point mass (SMBH) at the centre of the cluster as the orbital period of the most bound light particle (star) would then determine the running time of the simulation.

\subsection{Strong scaling tests}

We perform a series of strong scaling tests to study the overall scaling behaviour of our \mstar{} integrator. The results of the strong scaling test of the force calculation part of the code were presented in Fig. \ref{fig: forcescaling}. As before in a strong scaling test the problem size remains fixed while the number of CPUs is increased. In our six tests we use six different logarithmically spaced particle numbers with $264 \leq N_\mathrm{part} \leq 10^4$ as in Section \ref{Section: para-full}. The strong scaling tests consists of in total $270$ short N-body simulations with initial conditions described in the previous Section. In the simulations each of the point-mass clusters is propagated for $T = 0.1$ Myr which is close to the crossing times of the point-mass clusters. The GBS tolerance is set to $\etabs = 10^{-6}$ in these tests. We test CPU numbers up to $N_\mathrm{CPU}=400$. The CPU number $N_\mathrm{CPU}$ is always divided between the force loop tasks ($N_\mathrm{force}$) and substep parallelisation ($N_\mathrm{div}$) in a way which minimises the simulation running time as explained in Section \ref{Section: para-full}.

\begin{figure}
\includegraphics[width=\linewidth]{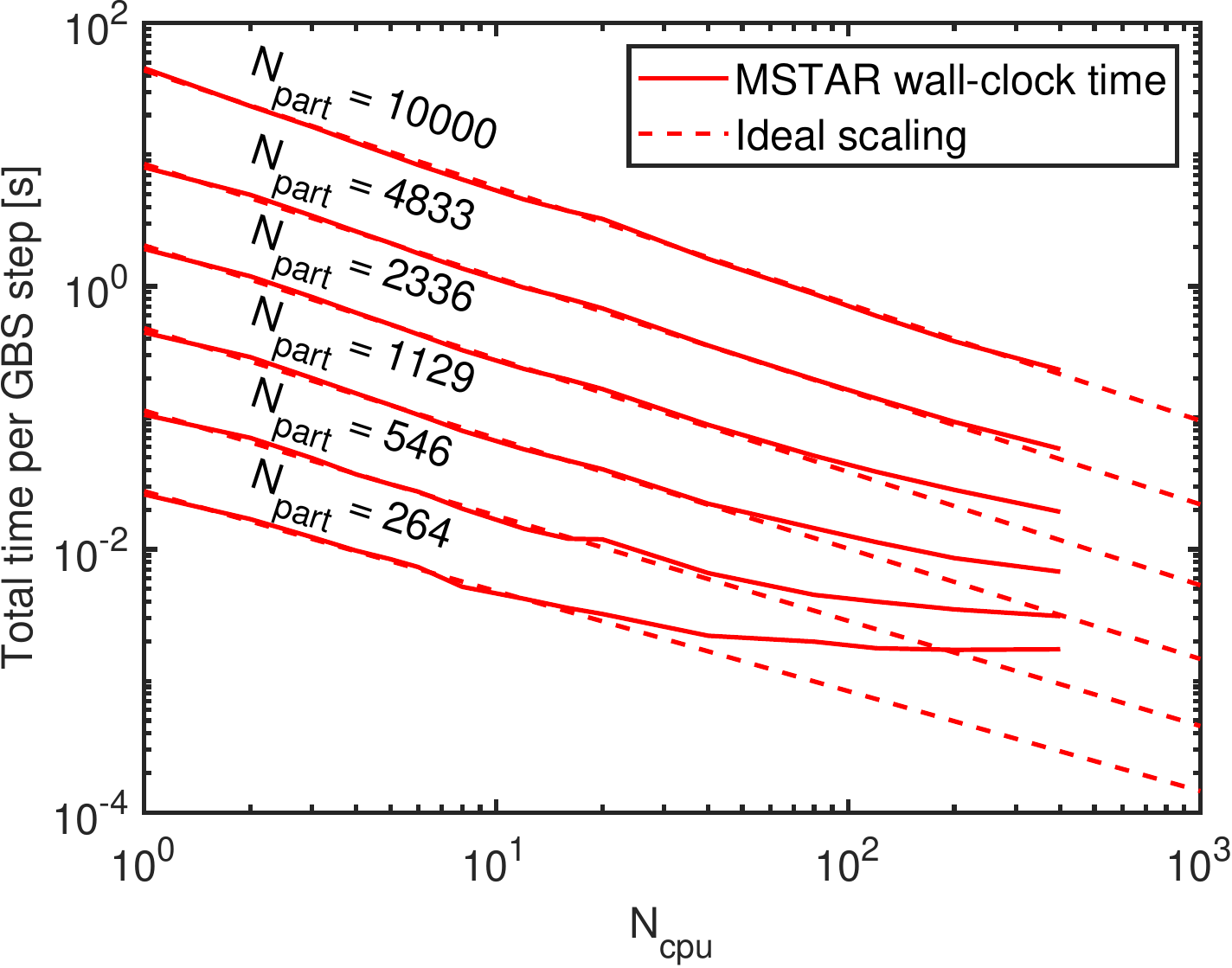}
\caption{The results of the strong scaling test of our \mstar{} integrator (solid red line). The ideal scaling behaviour is indicated by the dashed red line. The scaling behaviour of the integrator begins to deviate from the ideal scaling law around $N_\mathrm{CPU} \sim 0.1 \times N_\mathrm{part}$ and its completely saturated around $N_\mathrm{CPU} \sim 0.2 \times N_\mathrm{part}$. The test setups with the two highest particle numbers tested retain the ideal scaling behaviour up to the largest CPU number used in these tests, $N_\mathrm{CPU} = 400$.}
\label{fig: timing_wang2015like}
\end{figure}

The results of the strong scaling tests are shown in Fig. \ref{fig: timing_wang2015like}. The maximum speed-up factors with $N_\mathrm{CPU}=400$ range from $S_\mathrm{total} \approx 15$ with $N_\mathrm{part}=264$ to $S_\mathrm{total} \approx 145$ when $N_\mathrm{part}=10^4$. At this point the scaling of the setup with the lower particle number $N_\mathrm{part}$ is completely saturated while the setup with $N_\mathrm{part} = 10^4$ would still benefit from additional computational resources. However, we do not pursue numerical experiments beyond $N_\mathrm{CPU}=400$ in this study. We find that the integrator follows ideal scaling behaviour roughly up to the CPU number of $N_\mathrm{CPU} \sim 0.1 \times N_\mathrm{part}$ and a flat, saturated scaling occurs with $N_\mathrm{CPU} \gtrsim 0.2 \times N_\mathrm{part}$. The scaling of the entire code starts to deviate from the 
ideal scaling behaviour at a factor of a few smaller $N_\mathrm{CPU}$ than the scaling of only the force calculation part of the code. We attribute this difference to Amdahl's law \citep{Amdahl1967}, which states that maximum speed-up of a parallelised code depends on the fraction of serial code or code which cannot be efficiently parallelised. The force calculation part of the code can be almost entirely parallelised except for the necessary MPI communication between the CPUs. The entire integrator contains additional parts which cannot be parallelised as efficiently as the force computation. The main functions containing these serial code sections or functions which are difficult to parallelise efficiently are the \MST{} construction functions and the GBS extrapolation procedure.

\subsection{Timing tests}

We perform another set of N-body simulations to evaluate how much faster the parallelised version of the \mstar{} integrator is than our \archain{} implementation in \cite{Rantala2017}. In the simulations each of the $60$ point-mass clusters is again propagated for $T = 0.1$ Myr with a GBS tolerance of $\etabs = 10^{-6}$. The other parameters remain as in the previous Sections.

\begin{center}
\begin{table}
\begin{tabular}{ |l|c|c|c| } 
 \hline
 Label & Integrator & Mode & Resources \\ 
 \hline
 R17-S-1 & \archain{} & serial   & $1$ CPU \\ 
 R17-P-24 & \archain{} & parallel & $24$ CPU \\ 
 R20-S-1 & \mstar{} & serial   & $1$ CPU \\ 
 R20-P-max & \mstar{} & parallel & $2$-$400$ CPU \\ 
 \hline
\end{tabular}
\caption{The integrators and their serial/parallel configurations studied in the running time test. In the setup R20-P-max we selected the CPU number within $2\leq N_\mathrm{CPU} \leq 400$ for each particle number which yielded the fastest simulation times.}
\label{table: timing}
\end{table}
\end{center}
We test four different integrator configurations. The integrator details are collected in Table \ref{table: timing}. Our old \archain{} implementation is used both in a serial (R17-S-1) and a parallel mode (R17-P-24) with 24 CPUs as in \cite{Rantala2017}. We test the \mstar{} integrator in serial and parallel modes as well (R20-S-1 and R20-P-max). In the test setup R20-P-max we experimented with CPU numbers within $2\leq N_\mathrm{CPU} \leq 400$ and chose the $N_\mathrm{CPU}$ with gave the smallest running time for each particle number. In general adding more CPUs speeds up the computation until the scaling stalls around $N_\mathrm{CPU} \sim 0.2 \times N_\mathrm{part}$ as already illustrated in Fig. \ref{fig: timing_wang2015like}. This type of a test is not performed with our old \archain{} integrator as the scaling of the code becomes poor beyond a few tens of CPUs \citep{Rantala2017}.

\begin{figure}
\includegraphics[width=\linewidth]{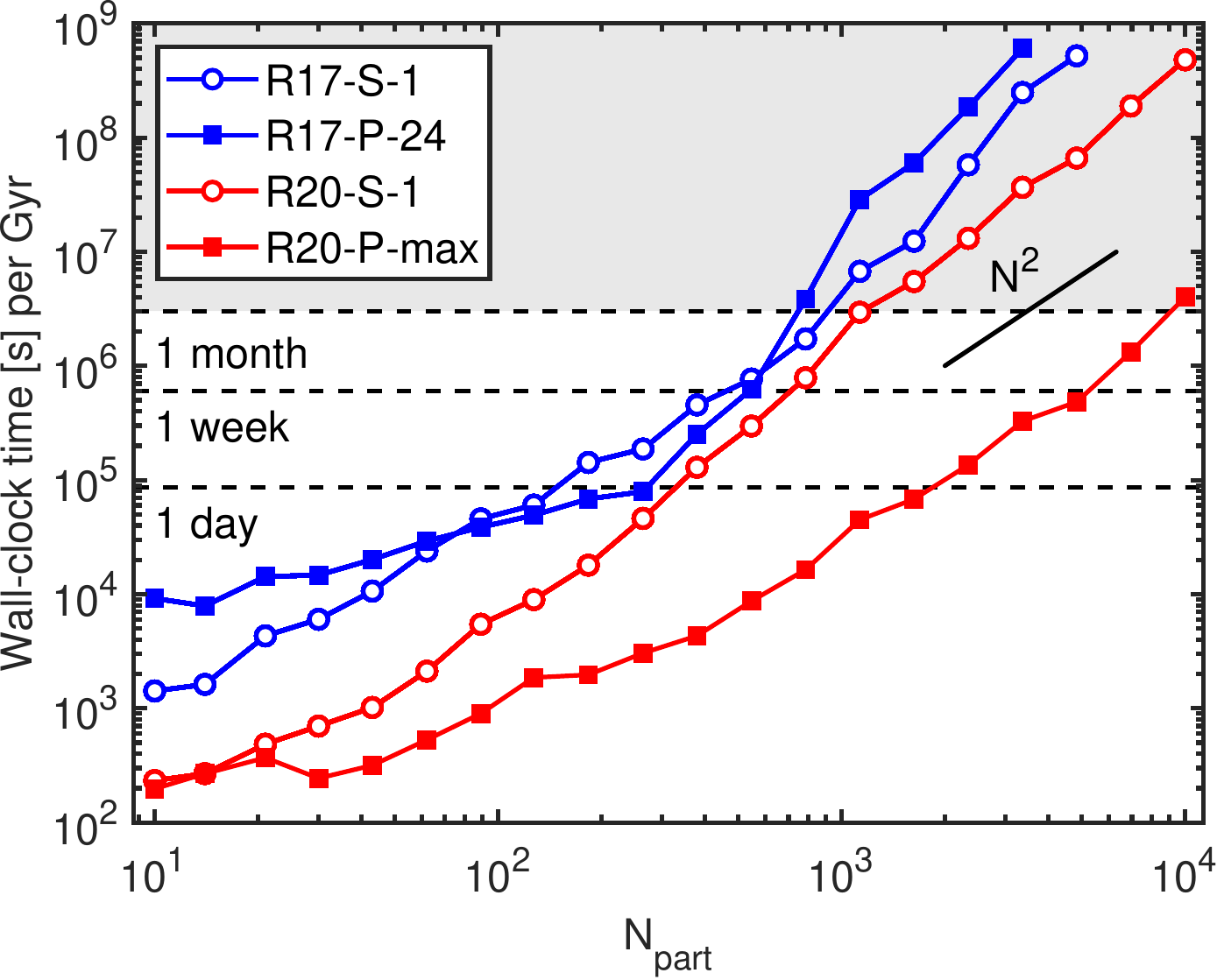}
\caption{The timing test of our \archain{} (blue) and \mstar{} (red) integrators. The parallel runs are indicated with filled squares while the serial runs are labelled with open circles. Benchmarks of $1$ day, $1$ week and $1$ month are also included. The grey shaded area marks the region which we deem too time-consuming for practical applications. We note that the new \mstar{} serial code is even faster than the old parallel \archain{} integrator and that the new parallel code is extremely fast compared to the other implementations, especially for large particle numbers.}
\label{fig: timing}
\end{figure}

The wall-clock times elapsed in the timing tests of the four integrator configurations are presented in Fig. \ref{fig: timing} as a function of the particle numbers of the simulated clusters. The results are scaled into the units of wall-clock time in seconds per simulated Gyr. We restrict our tests to simulations which last less than $10^9$ seconds of wall-clock time in the scaled units. Studying the results with the \archain{} integrator first, we see that the parallel implementation is faster than the serial version in the particle number range $50 \lesssim N_\mathrm{part} \lesssim 700$. Our simulations in previous studies \citep{Rantala2018,Rantala2019} have used regularised particle numbers in the lower half of this range. With lower particle numbers the communication time exceeds the speed-up from parallelised force loops. The slow-down of the parallelised old integrator at high particle numbers $N_\mathrm{part}\gtrsim 700$ is attributed to the fact that the code is only partially parallelised as serial functions still remain in the parallelised version.

Comparing the serial implementations of the \mstar{} and the \archain{} integrator (R17-S-1 and R20-S-1) we see that the new integrator is faster by a factor of $\sim 6$ when $N_\mathrm{part} \lesssim100$. The speed difference of the two serial codes reaches its minimum value of $\sim 2$ around $N_\mathrm{part}=10^3$. After this the speed difference of the codes begins to increase again reaching a value of $\sim 8$ at $N_\mathrm{part}=5 \times 10^3$. We note that the \mstar{} serial implementation R20-S-1 is even faster than the \archain{} parallel version R17-P-24 in every simulation we run for this study. Code run-time performance analysis tools reveal that the cache efficiency of our \archain{} code is poor compared to the new \mstar{} code. This fact explains the run-time speed difference of the two serial codes. All the four integrator configurations fairly closely follow the $\bigO(N_\mathrm{part}^2)$ scaling, being consistent with the theoretical scaling of the regularisation algorithm of $\bigO(N_\mathrm{part}^{2.13 - 2.20})$.

Studying the results of the \mstar{} integrator we can see that the parallel setup R20-P-max is always faster than the serial setup R20-S-1, even with small particle numbers. In addition, the test runs with R20-P-max become increasingly faster than the serial setup towards high particle numbers. Within $10^3 \lesssim N_\mathrm{part} \lesssim 10^4$ the speed-up factor is $\sim 55$-$145$. Compared to the fastest old \archain{} implementation the new parallel \mstar{} code is faster by a very large factor of $\sim 1100$ in this range of particle numbers.

The adopted GBS accuracy parameter affects the wall-clock time elapsed in the simulations. We perform an additional brief series of timing tests using \mstar{} with simulation parameters $N_\mathrm{part} = 1129$, $N_\mathrm{CPU} = 200$ and $10^{-12} \leq \etabs \leq 10^{-6}$. We find that in our tests the elapsed wall-clock time $T$ scales well as a power-law as
\begin{equation}
    \frac{T}{T(\etabs = 10^{-6})} = \left( \frac{\etabs}{10^{-6}} \right)^{-\alpha}
\end{equation}
in which the power-law index $\alpha \approx 0.05$ when $10^{-10} \lesssim \etabs \leq 10^{-6}$ and $\alpha \approx 1$ when $\etabs \lesssim 10^{-10}$. Due to the mild scaling of the wall-clock time $T$ as a function of the GBS accuracy parameter $\etabs$ we conclude that the results in the Section run with $\etabs = 10^{-6}$ apply in general for GBS tolerances $\etabs \gtrsim 10^{-10}$. However, we stress that the timing results of the codes depend on the N-body particle setups used, with more centrally concentrated stellar systems requiring in general more computational time.

Finally, we end this Section by discussing how large simulations can be run with our new \mstar{} code within a reasonable wall-clock time. We have marked the running times of $1$ day, $1$ week and $1$ month into Fig. \ref{fig: timing}. The grey area in the illustration marks the running times beyond $1$ month per Gyr which we consider unacceptably time-consuming. The parallel \mstar{} code can perform simulations with of the order of $10$ times more N-body particles with similar wall-clock times as our old integrator implementations. Simulations with $4000$-$7000$ particles are expected to last a few weeks with the parallel \mstar{} code with $N_\mathrm{CPU} = 400$. Running a simulation with $N_\mathrm{part} = 10^4$ in a similar wall-clock time would require $N_\mathrm{CPU} \approx 2000$ cores.

\begin{table*}
\begin{center}
\begin{tabular}{ |l|c|c|c|c| } 
 \hline
& \cite{Ito1997} & \cite{Mikkola2008} & \cite{Rantala2017} & This work\\ 
 \hline
 Regularisation & $\times$ & $\checkmark$ & $\checkmark$ & $\checkmark$\\ 
 Extrapolation method & $\checkmark$ & $\checkmark$ & $\checkmark$ & $\checkmark$\\ 
 \hline
 Chained coordinates &$\times$& $\checkmark$ & $\checkmark$ & $\checkmark$\\ 
 \MST{} coordinates     &$\times$& $\times$ & $\times$ & $\checkmark$\\ 
 \hline 
 Serial code         &$\checkmark$& $\checkmark$ & $\checkmark$ & $\checkmark$\\ 
 Parallel force loops      &$\times$& $\times$ & $\checkmark$ & $\checkmark$\\
 Parallel GBS subdivisions &$\checkmark$& $\times$ & $\times$ & $\checkmark$\\  
 \hline
\end{tabular}
\caption{A brief summary of the main properties of the \mstar{} integrator presented in this work alongside related integration methods from the literature.}
\label{table: parallel}
\end{center}
\end{table*}

\section{Conclusions} \label{Section: conclusions}

We have developed and implemented the \mstar{} integrator, a new fast algorithmically regularised integrator. While the time transformation scheme of the regularised integrator remains the same as in the earlier \archain{} integrator, the coordinate system and the GBS extrapolation method are significantly improved. A brief summary of the main ingredients of our integrator code and a selection of related integrators from the literature is collected in Table \ref{table: parallel}.

In our new \mstar{} implementation the chained coordinate system of \archain{} is replaced by a minimum spanning tree coordinate system, which can be viewed as a branching chain structure. Due to its ability to branch, the \MST{} avoids the possible pathological chain configurations in which spatially close particles can be found in very different parts of the chain. We find that the numerical error originating from building and deconstructing the coordinate structures is approximately smaller by a factor of $\sim 10$ for the \MST{} compared to the chain. The reason for this is that the \MST{} is a much shallower data structure than the chain as the average number of inter-particle vectors to reach the root vertex of the system is smaller. Thus, we recommend using the \MST{} coordinate system instead of the chained coordinates even though the code implementation becomes somewhat more complicated.

Our \mstar{} integrator includes a simplified GBS extrapolation method with two layers of MPI parallelisation. First, the force loop computation is parallelised with $N_\mathrm{force}$ CPUs with one MPI task each. The second layer is included in order to compute the $k_\mathrm{fix}$ substep divisions in parallel using $N_\mathrm{div}$ CPU groups. We also provide a recipe for estimating how to divide the computational resources to the different parallelisation layers and estimates for the maximum reasonable number of CPUs for parallelisation before the code scaling stalls.

We validate the numerical accuracy of our \mstar{} integrator in a series of demanding two- and three-body test simulations. The simulation setups include an eccentric Keplerian binary, the classic Pythagorean three-body problem and Lidov--Kozai oscillations in a hierarchical three-body system. Overall the particle orbits in the test runs are practically identical with both the \mstar{} and \archain{} integrators. In fact \mstar{} conserves energy, angular momentum and the direction of the Laplace--Runge--Lenz vector somewhat better than our previous regularised integrator \archain.

We test the speed and scaling behaviour of the \mstar{} integrator in a series of N-body stellar cluster simulations with up to $N_\mathrm{part} = 10^4$ particles. We find that the new integrator is always faster than the \archain. With full parallelisation we can efficiently use $\sim 10$ times more CPUs with adequate code scaling behaviour compared to our integrator implementation with force loop parallelisation only. The speed-up gained by the new fully parallelised integrator is substantial. The parallel \mstar{} code is up to a factor of $\sim 145$ faster than the serial \mstar{} code and up to a factor of $\sim 1100$ faster than the \archain{} code when the simulation particle number in the range $10^3 \lesssim N_\mathrm{part} \lesssim 10^4$.

The \mstar{} integrator will be important when pressing towards the ultimate goal of running collisionless simulations containing regularised regions with collisional stars and SMBHs with up to $N_\mathrm{part}\sim 5 \times 10^{8} \text{--} 10^{9}$ simulation particles in individual galaxies. We estimate that the \mstar{} integrator is able to run a Gyr-long simulation with $N_\mathrm{part} = 10^4$ in approximately two weeks of wall-clock time using $N_\mathrm{CPU} \approx 2000$ CPUs. In our previous studies with the \ketju{} code \citep{Rantala2017} which couples the \gadget{} tree code to the \archain{} integrator the total particle number in galaxies was limited to $N_\mathrm{part} \lesssim 10^7$ particles \citep{Rantala2018,Rantala2019}. This is due to the fact that our \archain{} integrator could efficiently handle only up to $200$-$300$ particles in the regularised regions. Based on these numbers we estimate that \mstar{} can treat $\sim 50$ times more regularised particles than our \archain{} implementation in a reasonable wall-clock time. Thus, galaxy simulations with accurate SMBH dynamics using \mstar{} in \ketju{} instead of \archain{} containing $5 \times 10^8 \lesssim N_\mathrm{part} \lesssim 10^9$ simulation particles seem achievable in the immediate future. These particle numbers yield stellar mass resolutions down to $m_\mathrm{\star} \approx 100 M_\odot$ even for simulations of massive early-type galaxies.

Finally, the improved numerical scaling and performance will be crucial when simulating the dynamics of SMBHs in gas-rich galaxies, which
are expected to have steep central stellar concentrations due to elevated levels of star formation in their nuclei. This is in particular important as the upcoming LISA gravitational wave observatory will be 
most sensitive for SMBHs with masses in the range of $M_{\rm BH} \sim 10^{6} \text{--} 10^{7} M_{\odot}$ \citep{2007CQGra..24R.113A}, which are expected to reside at the centres of gas-rich late-type galaxies.

\section*{Acknowledgements}
We would like to thank Seppo Mikkola, the referee of the paper. The numerical simulations were performed on facilities hosted by the CSC -- IT Center for Science, Finland and the Max Planck Computing and Data facility (MPCDF), Germany. A.R., P.P., M.M. and P.H.J. acknowledge the support by the European Research Council via ERC Consolidator Grant KETJU (no. 818930). TN acknowledges support from the Deutsche Forschungsgemeinschaft (DFG, German Research Foundation) under Germany's Excellence Strategy - EXC-2094 - 390783311 from the DFG Cluster of Excellence "ORIGINS".

\bibliographystyle{mnras}
\bibliography{references}

\bsp	
\label{lastpage}
\end{document}